\documentclass[traditabstract]{aa}

\usepackage{graphicx}
\usepackage{txfonts}
\usepackage{natbib}
\usepackage{amssymb}
\usepackage{amsfonts}
\usepackage{amsbsy}
\usepackage{amsmath}
\usepackage{pdflscape}

\newcommand{\Ab}{\boldsymbol{A}}
\newcommand{\Bb}{\boldsymbol{B}}
\newcommand{\fb}{\boldsymbol{f}}

\newcommand{\gb}{\boldsymbol{g}}
\newcommand{\Cb}{\boldsymbol{C}}

\newcommand{\hb}{\boldsymbol{h}}
\newcommand{\Ib}{\boldsymbol{I}}

\newcommand{\nb}{\boldsymbol{n}}

\newcommand{\ssb}{\boldsymbol{s}}
\newcommand{\rb}{\boldsymbol{r}}

\newcommand{\xb}{\boldsymbol{x}}

\newcommand{\yb}{\boldsymbol{y}}

\newcommand{\Tc}{\mathcal{T}}

\newcommand{\Smatb}{\boldsymbol{{\mathcal S}}}
\newcommand{\Xmatb}{\boldsymbol{{\mathcal X}}}

\newcommand{\Ematb}{\boldsymbol{{\mathcal E}}}

\newcommand{\Nmatb}{\boldsymbol{{\mathcal N}}}

\begin{document} 

\title{The correct estimate of the probability of false detection \\ of the matched filter  in weak-signal detection problems. II}
\subtitle{Further results with application to a set of ALMA and ATCA data}
 
   \author{R. Vio
          \inst{1}, C. Verg\`es \inst{2,3}
          \and
          P. Andreani 
         \inst{2}
         }

   \institute{Chip Computers Consulting s.r.l., Viale Don L.~Sturzo 82,
              S.Liberale di Marcon, 30020 Venice, Italy\\
              \email{robertovio@tin.it}
          \and
           ESO, Karl Schwarzschild strasse 2, 85748 Garching, Germany \\
             \email{pandrean@eso.org}             
         \and
         	\'{E}cole polytechnique, Route de Saclay, 91120 Palaiseau, France
				}
   \date{Received....; accepted....}

 \abstract{The matched filter (MF) is one of the most popular and reliable techniques to the detect signals of known structure and amplitude smaller than the level of the contaminating noise.
Under the assumption of stationary Gaussian noise, 
MF maximizes the probability of detection subject to a constant probability of false detection or false alarm (PFA). This property relies upon a priori knowledge of the position of the searched signals, which is usually not
available. Recently, it has been shown that when applied in its standard form, MF may severely underestimate the PFA.
As a consequence the statistical significance of features that belong to noise is overestimated and the resulting detections are actually spurious. For this reason, an alternative method of computing the 
PFA has been proposed that is based on the probability density function (PDF) of the peaks of an isotropic Gaussian random field. In this paper we further develop this method. In particular,  we discuss 
the statistical meaning of the PFA and show that, although 
useful as a preliminary step in a detection procedure,  it is not able to quantify the actual reliability of a specific detection. For this reason, a new quantity is introduced called the specific probability of false alarm (SPFA), which is able
to carry out this computation. We show how this method works in targeted simulations and apply it to a few interferometric maps taken with the Atacama Large Millimeter/submillimeter Array (ALMA) and the Australia Telescope Compact Array (ATCA). 
We select a few potential new point sources and assign an accurate detection reliability to these sources.}
   \titlerunning{A correct computation of the probability of false detection of the matched filter}
   \authorrunning{Vio \& Andreani}
   \maketitle
 
\section{Introduction}

In many research and engineering areas the matched filter (MF) represents  the standard tool for the detection of signals with amplitude smaller than the level of the contaminating noise \citep{kay98, tuz01, lev08, mac05}.
The MF is a linear filter that optimally filters out the Fourier frequencies where the noise is predominant, 
preserving those where the searched signal gives a greater contribution. In many situations, the MF displays the best performance, providing the greatest probability of true detection subject to a constant probability of false detection or false alarm
(PFA). The widespread success of the MF is due to its reliability and resulting robustness. The main limitation of this approach lies in the implicit assumption that the position of the signal within a sequence of data 
(e.g., an emission line in a spectrum) or on an image (e.g., a point source in an astronomical map) is known. In most practical applications this is not the case.
For this reason, the MF is used assuming that, if present, the position of a signal corresponds to a peak of the filtered data.
In a recent paper, \citet{vio16} have shown that, when based on the standard but wrong assumption that the probability density function (PDF) of the peaks of a Gaussian noise process is a Gaussian, this approach may lead to a severe underestimation of the PFA. The same authors have provided an alternative procedure to correctly compute this quantity. 

The PFA is a useful tool to fix a preliminary detection threshold, but with this quantity alone it is not possible to assess the reliability of a specific detection. This is because it does not provide the probability that a given detection is spurious but rather 
the probability that a generic peak due to the noise in the filtered signal can exceed, by chance, a fixed detection threshold. However, there are often cases in which the knowledge of the probability for a peak to be a source is crucial to plan follow-up observations to identify the source itself. For this reason, in this work we introduce the so-called specific probability of false alarm (SPFA), which is able to provide a precise (within a few percent) quantification of the reliability of a detection.

In Sect.~\ref{sec:MF1}, the main characteristics of MF for one-dimensional signals are reviewed as well as the reason why it underestimates the PFA. In the same section the method suggested by \citet{vio16} 
to correctly compute this quantity is briefly reconsidered.
In Sect.~\ref{sec:spfa} we address the question of  the statistical meaning of the PFA and introduce the SPFA.
The arguments are extended to the two-dimensional signals in Sect.~\ref{sec:twodimensional}. Finally,  in Sect.~\ref{sec:experiment} the procedure is applied to a simulated map, whereas in Sects.~\ref{sec:real}  and ~\ref{sec:atca} it is applied
to two interferometric maps obtained with the Atacama Large Millimeter/submillimeter Array (ALMA) and to one map obtained with the Australia Telescope Compact Array (ATCA). The final remarks are deferred to Sect.~\ref{sec:conclusions}.

\section{Matched filter for one-dimensional signals} \label{sec:MF1}

\subsection{The basics} \label{sec:basicMF}

Given a discrete observed signal $\xb =  [x(0), x(1), \ldots, x(N-1)]^T$ of length $N$, the model assumed in the MF approach is
$\xb = \ssb + \nb$, where $\ssb$ is the deterministic signal to detect and $\nb$ a zero-mean Gaussian stationary noise  with known covariance matrix
\begin{equation} \label{eq:C}
\Cb = {\rm E}[\nb \nb^T]. 
\end{equation}
Here, symbols  ${\rm E}[.]$ and  $^T$ denote the expectation operator and the vector or matrix transpose, respectively.

Under these conditions,  according to the Neyman-Pearson theorem \citep{kay98,vio16}, a detection is claimed when
\begin{equation} \label{eq:test1}
\Tc(\xb) = \xb^T \fb > \gamma,
\end{equation}
with
\begin{equation} \label{eq:mf}
\fb_s = \Cb^{-1} \ssb.
\end{equation}
Here $\fb_s$, an $N \times 1$ array,  represents the matched filter. The main characteristic of the MF is that it maximizes of the probability of detection (PD) under the constraint of a fixed PFA.

\subsection{Matched filter in practical applications} \label{sec:comments}

The MF  has been obtained under two assumptions: first, signal $\ssb$ is known and, second, $\xb$ and $\ssb$ have the same length $N$. This last point implicitly means that the position of $\ssb$ within 
$\xb$  is known. 

Very often only the shape $\gb$ of the signal $\ssb =a \gb$ is known but not its amplitude $a$.
In this case, the relaxation of the first assumption does not have important consequences given that the test in Eq.~(\ref{eq:test1}) can be rewritten in the form
\begin{equation} \label{eq:test2}
\Tc(\xb) = \xb^T \fb_g > \gamma',
\end{equation}
where $\gamma'=\gamma/a$ and the MF becomes
\begin{equation} \label{eq:mf2}
\fb_g = \Cb^{-1} \gb.
\end{equation}
Hence, the resulting  $\Tc(\xb)$ is a statistic that is independent of $a$. This means that the unavailability of $a$ does not affect the PFA but only the PD.

If the second assumption is loosened, the signal  $\ssb$ has a length $N_s$ that is smaller than the length of the observed data (e.g. an emission line in an experimental spectrum). 
If the amplitude $a$ is also unknown, the standard approach is to claim a detection when
\begin{equation} \label{eq:test2b}
\Tc(\xb, \hat{i}_0) = \xb^T \fb_g > u \hat{\sigma}_{\Tc},
\end{equation}
where
\begin{equation} \label{eq:dectx}
\Tc(\xb, \hat{i}_0) = \underset{ i_0 \in [0, N-N_s] }{\max} \Tc(\xb, i_0),
\end{equation}
where $\hat{i}_0$ is the estimate of the unknown position $i_0$ of the signal, $u$ a value typically in the range $[3, 5]$, and $\hat{\sigma}_{\Tc}$ the standard deviation of the sequence
\begin{equation} \label{eq:corrx}
 \Tc(\xb, i_0) = \sum_{i=i_0}^{i_0 + N_s - 1} x(i) f_g(i-i_0); \quad i_0 = 0, 1, \ldots, N-N_s.
\end{equation}
Namely, the observed  signal $\xb$ is cross-correlated with the MF~\eqref{eq:mf2}; see Eq.~\eqref{eq:corrx}.
The greatest peak detected in $ \Tc(\xb, i_0) $; see Eq.~\eqref{eq:dectx}. Finally this latter is tested if it exceeds a threshold set to $u$ times the standard deviation of the sequence $\Tc(\xb, i_0)$; see Eq.~\eqref{eq:test2b}.
In the affirmative case the peak corresponds to a detection, otherwise it is assumed to be due to the noise.
If the number of signals $\ssb$ present in $\xb$ is also unknown, this procedure has to be applied to all the peaks in $ \Tc(\xb, i_0)$.

It is widespread practice that the PFA corresponding to the statistics~\eqref{eq:test2b} is given by
\begin{equation} \label{eq:fd1}
\alpha = \Phi_c(u),
\end{equation}
where  $\Phi_c(u) = 1 - \Phi(u)$ and $\Phi(u)$ is the standard Gaussian cumulative distribution function \footnote{In \citet{vio16} $\Phi_c(.)$ is denoted as  $\Phi(.)$.}.
However, as shown by \citet{vio16}, such practice can lead to underestimate this quantity severely. The point is that the PDF of the peaks of a stationary Gaussian random signal is not a Gaussian as
implicitly assumed in Eq.~\eqref{eq:fd1}. For this reason, the correct PFA has to be estimated by means of
\begin{equation} \label{eq:corra}
 \alpha = \Psi_c(u),
\end{equation}
where 
\begin{equation}
\Psi_c(u)= 1 - \Psi(u)
\end{equation} 
with
\begin{equation}
\Psi(u)=\int_{-\infty}^{u} \psi(z) dz,
\end{equation} 
and
\begin{equation} \label{eq:pdf_z1}
\psi(z) = \frac{\sqrt{3 - \kappa^2}}{\sqrt{6 \pi}} {\rm e}^{-\frac{3 z^2}{2(3 - \kappa^2)}} + \frac{2 \kappa z \sqrt{\pi}}{\sqrt{6}} \phi(z) \Phi\left(\frac{\kappa z}{\sqrt{3 - \kappa^2}} \right)
\end{equation}
providing the PDF  \footnote{In \citet{vio16} $\Psi_c(.)$ is denoted as  $\Psi(.)$ and in Eqs.~(24)-(25) the function $\Phi(.)$ has to be intended as the Gaussian cumulative distribution function and not its complementary function 
as it erroneously appears.}
of the local maxima of a zero-mean, unit-variance, smooth stationary one-dimensional Gaussian random field  \citep{che15a, che15b}. Here,
\begin{equation} \label{eq:kd}
\kappa = - \frac{\rho'(0)}{\sqrt{\rho''(0)}},
\end{equation}
where $\rho'(0)$ and $\rho''(0)$ are, respectively, the first and second derivative with respect to $r^2$ of the two-point correlation function $\rho(r)$ at $r=0$, where $r$ is the inter-point distance. 
When  $\kappa=1$, the functional form of the two-point correlation function is a Gaussian. The condition of smoothness for the random
fields requires that $\rho(r)$ be differentiable at least six times with respect to $r$. In a recent work, \citet{che16} have shown that, contrary to what is claimed in  
\citet{che15a, che15b} and in \citet{vio16}, the constraint $\kappa \leq 1$ actually is not necessary for the validity of the PDF~\eqref{eq:pdf_z1}. 

The main problem in using Eq.~\eqref{eq:corra} is the estimation of parameter $\kappa$. One possibility, suggested by \citet{vio16}, is to obtain such a quantity from the fit of the discrete sample two-point correlation function of 
$\Tc(\xb, i_0)$ with an appropriate analytical function. The reason is that the estimation of the correlation function of the noise is also required by the MF and, therefore, it is not an additional condition of the procedure.
However, a reliable estimate of $\rho'(0)$ and $\rho''(0)$ is a very delicate issue. A robust alternative is to estimate $\kappa$ through a maximum likelihood approach
\begin{equation} \label{eq:ml}
\hat{\kappa} = \underset{\kappa }{\arg\max} \sum_{i=1}^{N_p} \log{\left(\psi(z_i; \kappa)\right)},
\end{equation}
where $\{ z_i \}$, $i=1,2,\ldots, N_p$, are the local maxima of $\Tc(\xb, i_0)$. This approach cannot be adopted if the number and/or  the amplitude of the point sources modify the statistical characteristics
of the noise background. The latter case is simple to deal with since it is sufficient to mask all the sources that clearly stand out from the noise. Conversely, in the first case the approach is unfeasible. A comparison of the histograms $H(x)$ of
the pixel values and $H(z)$ of the peak amplitudes with the respective PDFs $\phi(x)$ and $\psi(z)$ permits us to check if this condition is fulfilled (see more in Sects.~\ref{sec:experiment}-\ref{sec:atca}).

Once $\hat{\kappa}$ and $N_p$ are fixed, an additional benefit of the maximum likelihood approach is that 
by means of  Eq.~\eqref{eq:kd} and of the expected number $N^*_p$ of peaks per unit length \citep{che15b},
\begin{equation}  \label{eq:np}
N^*_p= \frac{\sqrt{6}}{2 \pi} \sqrt{- \frac{\rho''(0)}{\rho'(0)}},
\end{equation} 
it is possible to obtain an estimate of $\rho'(0)$ and $\rho''(0)$ as, 
\begin{align}
\rho'(0) & =  \frac{2}{3} \pi^2 (N^*_p)^2 \hat{\kappa}^2; \\
\rho''(0) & = \frac{4}{9} \pi^4 (N^*_p)^4 \hat{\kappa}^2.
\end{align}
In turn, by means of the Taylor expansion,
\begin{equation}
\rho(r) = 1 + r^2 \rho'(0) + \frac{1}{2} r^4 \rho''(0),
\end{equation}
it is possible to evaluate the form of $\rho(r)$ for $r \in (0, 1)$, an interval that in the case of discrete signals where $r$ can take only integer values is not computable by means of the classical estimators of the correlation function.

\subsection{Morphological analysis of the peaks} \label{sec:morph}

It is a common believe that it is possible to improve the rejection of spurious detections by means of a morphological analysis of the shape of the peaks in the filtered map $\Tc(\xb, i_0)$. Such a conviction is based on the assumption  that a peak
due to a signal $\ssb$ looks different from a peak due to noise. Unfortunately, the situation is more complex. 

Let us assume, for the moment, that $\nb$ is a standard Gaussian white noise. In the case of no signal, $\Tc(\xb, i_0)$ is obtained through the correlation of $\nb$ with $\fb=\gb$. As a consequence, it is a Gaussian stochastic process with 
autocorrelation function 
$\rho(\tau)$ given by 
\begin{equation}
\rho(\tau) = \gb_{\star},
\end{equation}
where $\gb_{\star} = \gb \otimes \gb$ with symbol $\otimes$ denoting the correlation operator. Here, the point is that $ \gb_{\star}$ is also the shape of the template $\gb$ after the matched filtering. Moreover, the conditional expectation of 
$\Tc(\xb, i_p +\tau)$ given $x[i_p]$, with $i_p$ the position of the peak, is
\begin{equation}
{\rm E}[\Tc(i_p+\tau~ |~ x[i_p])] = a \rho(\tau),
\end{equation}
where $a = \Tc(\xb, i_p)$ is the peak value. This means that in $\Tc(\xb, i_0)$ the expected shape of a peak due to the noise is identical to that of the signal $\ssb$ after the matched filtering. 

Something similar holds when the noise is of non-white type. Indeed, in the case of no signal,  the quantity $\Tc(\xb) = \xb^T  \Cb^{-1} \gb$ in  Eq.~\eqref{eq:test2} can be written in the form $\Tc(\xb) = \yb^T \hb$ where $\yb^T = \xb^T \Cb^{-1/2}$ and
$\hb = \Cb^{-1/2} \gb$. Now, ${\rm E}[\yb \yb^T] = {\rm E}[\Cb^{-1/2} \xb \xb^T \Cb^{-1/2}] = \Cb^{-1/2} {\rm E}[\xb \xb^T] \Cb^{-1/2} = \Cb^{-1/2} \Cb \Cb^{-1/2} = \Ib$. This means that the non-white case can be brought back to a white case
where now the matched filter takes the form $\fb = \hb$.

\subsection{Previous similar works on this issue}

To the best of our knowledge to date, the beforehand detection procedure is the first that makes use of  the statistical characteristics of the peaks of a Gaussian noise. In the past, \citet{lop05a, lop05b} have proposed two 
modifications of the MF, known as the scale-adaptive filter (SAF) and the biparametric scale-adaptive filter (BSAF), which are claimed to be able to outperform 
the MF in the sense of maximizing  the probability of detection subject to a constant probability of false  detection.
Contrary to the MF, these filters work not only with the amplitude of the  peaks but also with the corresponding curvature. However, such a claim is incompatible with the Neyman-Pearson theorem given that,  as seen in Sect.~\ref{sec:basicMF},
in the case of stationary Gaussian noise and conditioned on the a priori knowledge of the true position of the signal, the filter that maximizes  the probability of detection subject to a constant probability of false  detection is the MF. This means that, under the same conditions, no other filter can outperform it.
A careful reading of the above-mentioned works permits to realize that the optimal properties of SAF and BSAF hold only under an additional condition with respect to the MF, namely that the true position of the source coincides with a peak of the filtered noise. It is worth noting that this is an unrealistic condition. In fact, since the arguments supporting SAF and BSAF are developed in the framework of continuous signals, such a coincidence represents an event with probability zero. 
As a consequence, the plain and uncritical extension of SAF and BSAF to the case of discrete signals operated by \citet{lop05a, lop05b} is questionable. The same also applies to their numerical simulations since the comparison of the performance of the SAF, BSAF, and MF should have been based on all the sources present in the simulated signals and not only on the subset of the sources whose true position, after the filtering, coincide with that of an observed peak.

\section{Specific probability of false alarm (SFPA)} \label{sec:spfa}

Contrary to what one could believe at first glance, the PFA given by Eq.~\eqref{eq:corra} does not provide the probability $\alpha$ that a specific detection is spurious but the probability that a generic peak due to the noise in $\Tc(\xb, i_0)$  
can exceed, by chance, the threshold $u$.
If $N_p$ peaks due to the noise are present in $ \Tc(\xb, i_0)$, then a number  $\alpha \times N_p$  among them  is expected to exceed the prefixed detection threshold. 
For example, if in $ \Tc(\xb, i_0)$ there are $1000$ peaks, then there is a high probability that a detection with a PFA equal to $10^{-3}$ is spurious.
As a consequence, in spite of the low PFA, the reliability of the detection is actually small.
A possible strategy  to avoid this problem is to fix a threshold $u$ such as $\alpha \times N_p \ll 1$. 
However, in this way there is the concrete risk to be too conservative and miss some true detections. In literature some procedures are available to alleviate this kind of problem. They are essentially of non-parametric type (i.e. they are not able to exploit all the available information). A popular approach is represented by the false discovery rate (FDR) \citep{mil01, hop02}. Here, we propose a parametric solution that consists in a 
preselection based on the PFA and then in the computation of  the probability of false detection for each specific detection.  We call this specific probability of false alarm (SPFA).
This quantity can be computed by means of the order statistics, in particular by exploiting  the statistical characteristics of the greatest value of a finite sample of {\it identical and independently distributed} (iid) random variable from a given 
PDF  \citep{hog13}. Under the iid condition, the PDF $g(z_{\max})$ of the largest value among a set of $N_p$ peaks $\{ z_i \}$ is given by
\begin{equation} \label{eq:gz}
g(z_{\max}) = N_p \left[ \Psi(z_{\max}) \right]^{N_p-1} \psi(z_{\max}).
\end{equation}
Hence, the SPFA can be evaluated by means of
\begin{equation} \label{eq:intz}
\alpha = \int_{z_{\max}}^{\infty} g(z') dz'.
\end{equation}
Actually, since the peaks  of a generic isotropic Gaussian random field tend to cluster, the iid condition is not necessarily valid. However,  in situations where the two-point correlation function $\rho(r)$ of the noise is narrow with respect the
area spanned by the data (a basic situation for the application of the MF), this condition can be expected to hold with good accuracy. The rationale is that two points with a distance $r$ such that $\rho(r) \approx 0$ 
are essentially independent. The same
holds for two generic peaks. Hence, most of the peaks can be expected to be approximately  iid and  Eq.~\eqref{eq:gz} is still applicable but possibly with an effective number $N^+_p < N_p$ \citep[see Sect. 6 in][]{may05}.
This last point is due to the dependence among a set of random variable which lowers its number of degrees of freedom \footnote{The term degrees of freedom refers to the number of
items that can be freely varied in calculating a statistic without violating any constraints.}.

A way to measure the degree of dependence of the peaks is the two-point correlation function $\rho_p[d]$. This discrete function is computed on a 
set of non-overlapping and contiguous distance bins of size $\Delta d$
\begin{equation}
\rho_p[d]= \sum_{\substack{i,j = 1  \\ d - \Delta d/ 2 < t_i - t_j \le  d + \Delta d/ 2 }}^{N_d} \frac{z[t_i] z[t_j]}{N_d} / \sum_{i=1}^{N_p} \frac{z[t_i] z[t_i]}{N_p},
\end{equation}
with $N_d$ the number of peak pairs with a distance within the range $(d - \Delta d/ 2, d + \Delta d/ 2]$. It measures the tendency of two peaks with similar value to be next to each other.

The numerical evaluation of integral~\eqref{eq:intz} does not present particular difficulties since
\begin{align}
\alpha & = N_p \int_{z_{\max}}^{\infty} \left[\Psi(z) \right]^{N_p-1} d\Psi(z); \\
          & = \left[ \Psi(z) \right]^{N_p} \Big|_{z_{\rm max}}^{\infty}; \\
          & = 1-\left[ \Psi(z_{\rm max}) \right]^{N_p}.
\end{align}

If the number of signals present in  $\Tc(\xb, i_0)$ is unknown, the above procedure can be applied, in order of decreasing amplitude, to all peaks with a PFA smaller 
than a prefixed $\alpha$, and reducing $N_p$ by one unit after any confirmed detection. The last step is based on the rationale that if a peak can be assigned to a signal $\ssb$ in $\xb$, it can be removed from the set of the peaks related to the noise.

\begin{figure*}
        \resizebox{\hsize}{!}{\includegraphics{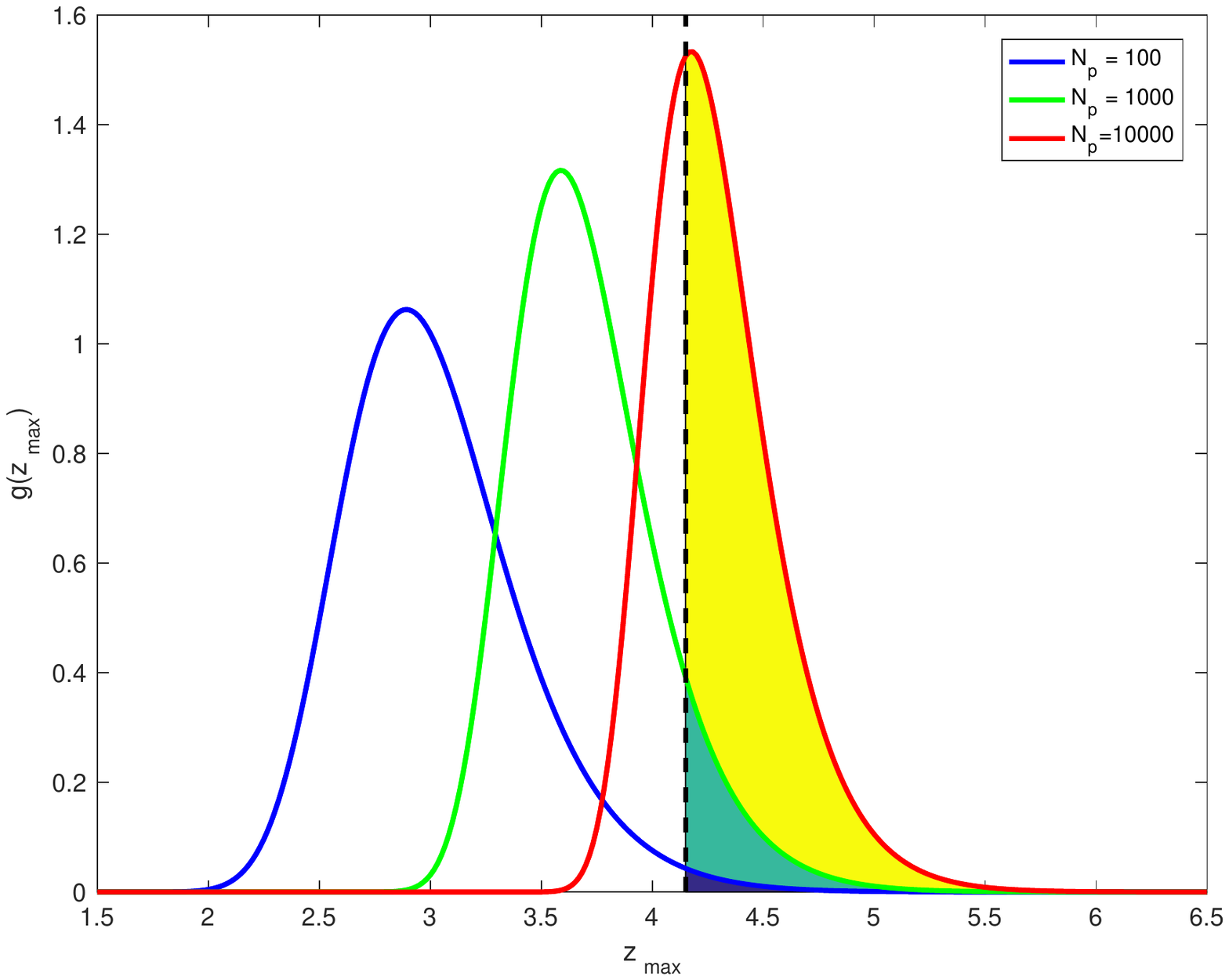}}
        \caption{PDF $g(z_{\rm max})$ of the greatest value of a finite sample of $N_p=10^2, 10^3$, and $10^4$ identical and independently distributed random variable from the PDF $\psi(z)$ of the peaks of a one-dimensional zero-mean unit-variance stationary Gaussian process field with $\kappa=1$. The color-filled areas provide the respective SPFA  for a detection threshold $u$ (dashed line) corresponding to a PFA~\eqref{eq:corra} equal to $10^{-4}$. It is evident that a detection threshold independent of $N_p$ is not able to quantify the risk of a false detection.}
        \label{fig01}
    \end{figure*}

The importance of the SPFA is demonstrated by Fig.~\ref{fig01} where the PDF $g(z_{\max})$,
corresponding to the PDF $\psi(z)$ of the peaks of a stationary zero-mean unit-variance Gaussian random process with $\kappa=1$, is plotted for three different values of the sample size $N_p$,
i.e., $10^2$, $10^3$, and $10^4$. The color-filled areas provide the respective SPFA  for a detection threshold $u$ corresponding to a PFA equal to $10^{-4}$.
It is evident that a detection threshold independent of 
$N_p$ is not able to quantify the risk of a false detection. This figure also shows that
the determination of the number $N^+_p$ is not a critical operation since $g(z_{\max})$ is a slow changing function of $N_p$ and for weakly dependent peaks it is $N_p \approx N_p^+$.
The SPFA is also useful because, by means of the Poisson-binomial distribution \footnote{We recall that the Poisson-binomial distribution is the probability distribution of the number of successes 
in a sequence of $N$ independent experiments having only two possible outcomes (yes/no) with success probabilities $p_1, p_2, \ldots, p_N$. When $p_1 = p_2 = \ldots = p_N$, it coincides with the binomial distribution.}, 
it is possible to estimate the probability that the number of false detection $N_{\rm FD}$  is equal or less than a given integer $k$ (see below).

\section{Extension of matched filter to the two-dimensional case} \label{sec:twodimensional}

In principle, the extension of MF to the two-dimensional signals $\Xmatb$, $\Smatb$, and $\Nmatb$ does not present particular difficulties. Indeed, it is sufficient to set
\begin{align}
\ssb & = {\rm VEC}[\Smatb]; \label{eq:stack1} \\
\xb & = {\rm VEC}[\Xmatb]; \label{eq:stack2} \\
\nb & = {\rm VEC}[\Nmatb], \label{eq:stack3}
\end{align}
where ${\rm VEC}[\Ematb]$ is the operator that transforms a matrix $\Ematb$ into a column array by stacking its columns one underneath the other, 
to obtain a problem similar to that in Sect.~\ref{sec:MF1}. There are only two differences. 
The first is the structure of matrix $\Cb$. In the one-dimensional case matrix $\Cb$ is of Toeplitz type, i.e., a matrix in which each descending diagonal from left to right is constant. In the two-dimensional case, however,
it becomes a block-Toeplitz with Toeplitz-block (BTTB) type, i.e., a matrix that contains Toeplitz blocks that are repeated down the diagonals of the matrix, as a Toeplitz matrix has elements repeated down the diagonal.
The second one concerns the PDF of the local maxima and their expected number per unit area that, in the case of a zero-mean unit-variance Gaussian isotropic noise, are given by \citep{che15a, che15b}
\begin{multline} \label{eq:pdf_z2}
\psi(z) = \sqrt{3} \kappa^2 (z^2-1) \phi(z) \Phi \left( \frac{\kappa z}{\sqrt{2 - \kappa^2}} \right) + \frac{\kappa z \sqrt{3 ( 2 - \kappa^2)}}{2 \pi} {\rm e}^{-\frac{z^2}{2 - \kappa^2}}\\
+\frac{\sqrt{6}}{\sqrt{\pi (3 - \kappa^2)}} {\rm{e}^{-\frac{3 z^2}{2 (3-\kappa^2)}}} \Phi\left( \frac{\kappa z}{\sqrt{(3 - \kappa^2) (2 - \kappa^2)}} \right).
\end{multline}
and
\begin{equation} \label{eq:number}
N^*_p = -\frac{\rho''(0)}{\pi \sqrt{3} \rho'(0)},
\end{equation} 
respectively. However, a computational problem arises because, even for maps of moderate size, the covariance matrix $\Cb$
becomes rapidly huge. Hence, some efficient numerical methods based on a Fourier approach have to be used as in
\citet[ ][Chap.~5]{vog02},  \citet[][page $145$]{jai89}, \citet[][Appendix B]{els13}, and \citet[][Appendix A]{lag91}.

\begin{figure*}
        \resizebox{\hsize}{!}{\includegraphics{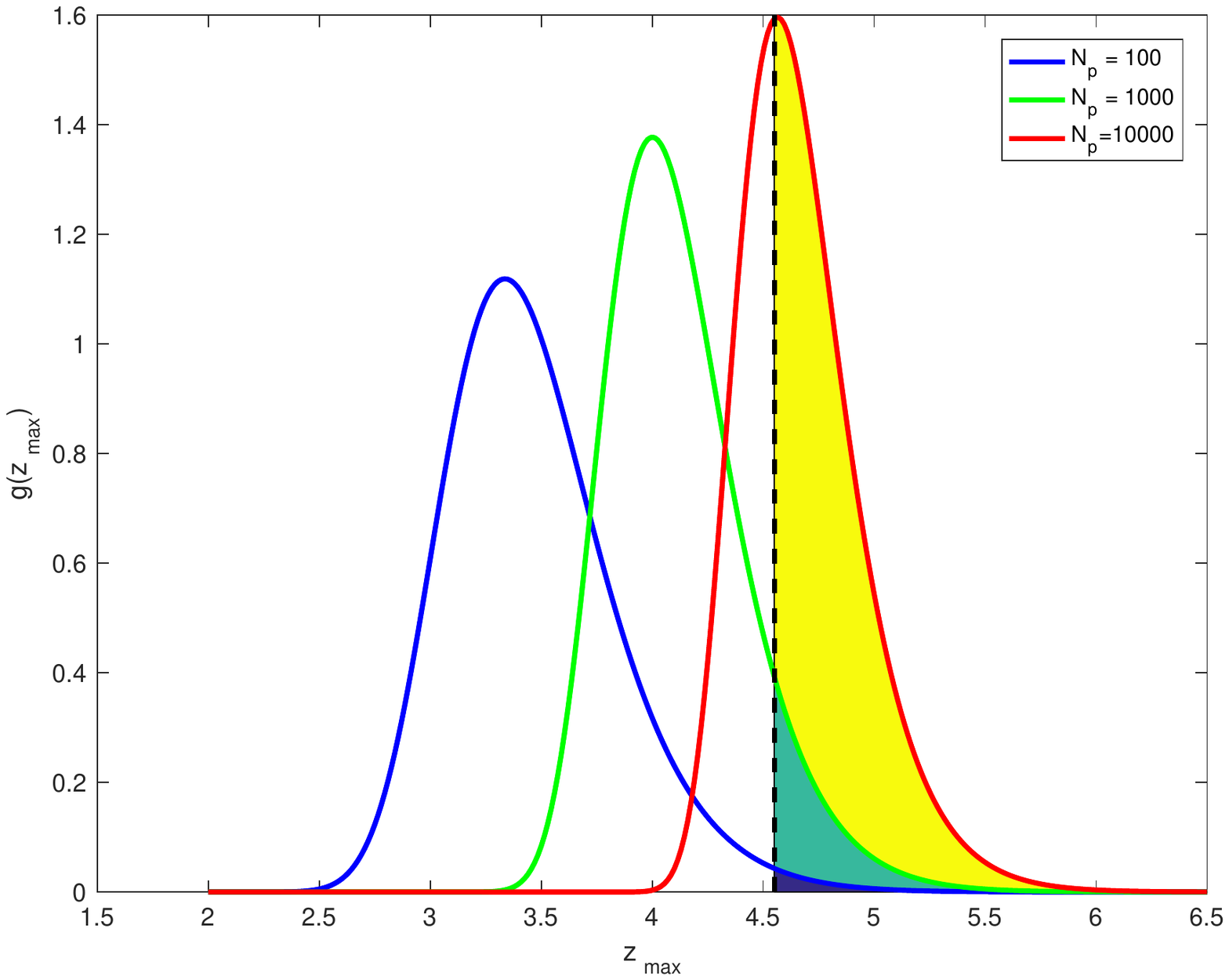}}
        \caption{PDF $g(z_{\rm max})$ of the greatest value of a finite sample of $N_p=10^2, 10^3$, and $10^4$ identical and independently distributed random variable from the PDF $\psi(z)$ of the peaks of a two-dimensional zero-mean unit-variance isotropic Gaussian random field with $\kappa=1$. The color-filled areas provide the respective SPFA  for a detection threshold $u$  (dashed line) corresponding to a PFA~\eqref{eq:corra} equal to $10^{-4}$. It is evident that a detection threshold independent of $N_p$ is not able to quantify the risk of a false detection.}
        \label{fig02}
    \end{figure*}

In comparison with Fig.~\ref{fig01}, Fig.~\ref{fig02} also shows the importance of the SPFA for the two-dimensional case.  
Here, the PDF $g(z_{\max})$, corresponding to the PDF $\psi(z)$ of the peaks of an isotropic two-dimensional zero-mean unit-variance Gaussian random field with $\kappa=1$, is plotted for three different values of the sample size $N_p$, $10^2$, $10^3$, and $10^4$.  Again, the color-filled areas provide the respective SPFA  for a detection threshold $u$ corresponding to a PFA equal to $10^{-4}$.

A final note concerns the condition of isotropy of the noise. There are situations where such condition can be relaxed. In particular this happens with correlation functions of type
\begin{equation}
\rho(r)= f(\rb^T \Ab \rb),
\end{equation} 
where $f(.)$ is a real function and $\Ab = \Bb^T \Bb$ with $\Bb$ a non-degenerated matrix. The random fields corresponding to this correlation function are not isotropic but only homogeneous (i.e., the statistical characteristics are constant along a given
direction). The arguments presented above can also be applied to this case given that an opportune rotation and rescaling of the axes can convert $\rb^T \Ab \rb$ into $\rb^T \rb$. Since these operations on the coordinate system 
do not modify the values of the random field, the PDF of the peaks does not change
\footnote{The rotation and the rescaling of the coordinate system change only the number of peaks for unit area, which becomes $|{\rm Det}[ \Bb  ]|^{1/2}$  times that of the isotropic case $\rho(\rb) = f(\rb^T \rb)$ (Cheng,
private communication). Here ${\rm Det}[.]$ denotes the determinant of a matrix.}.

\section{A numerical experiment} \label{sec:experiment}

\begin{landscape}
   \begin{figure}
        \resizebox{\hsize}{!}{\includegraphics{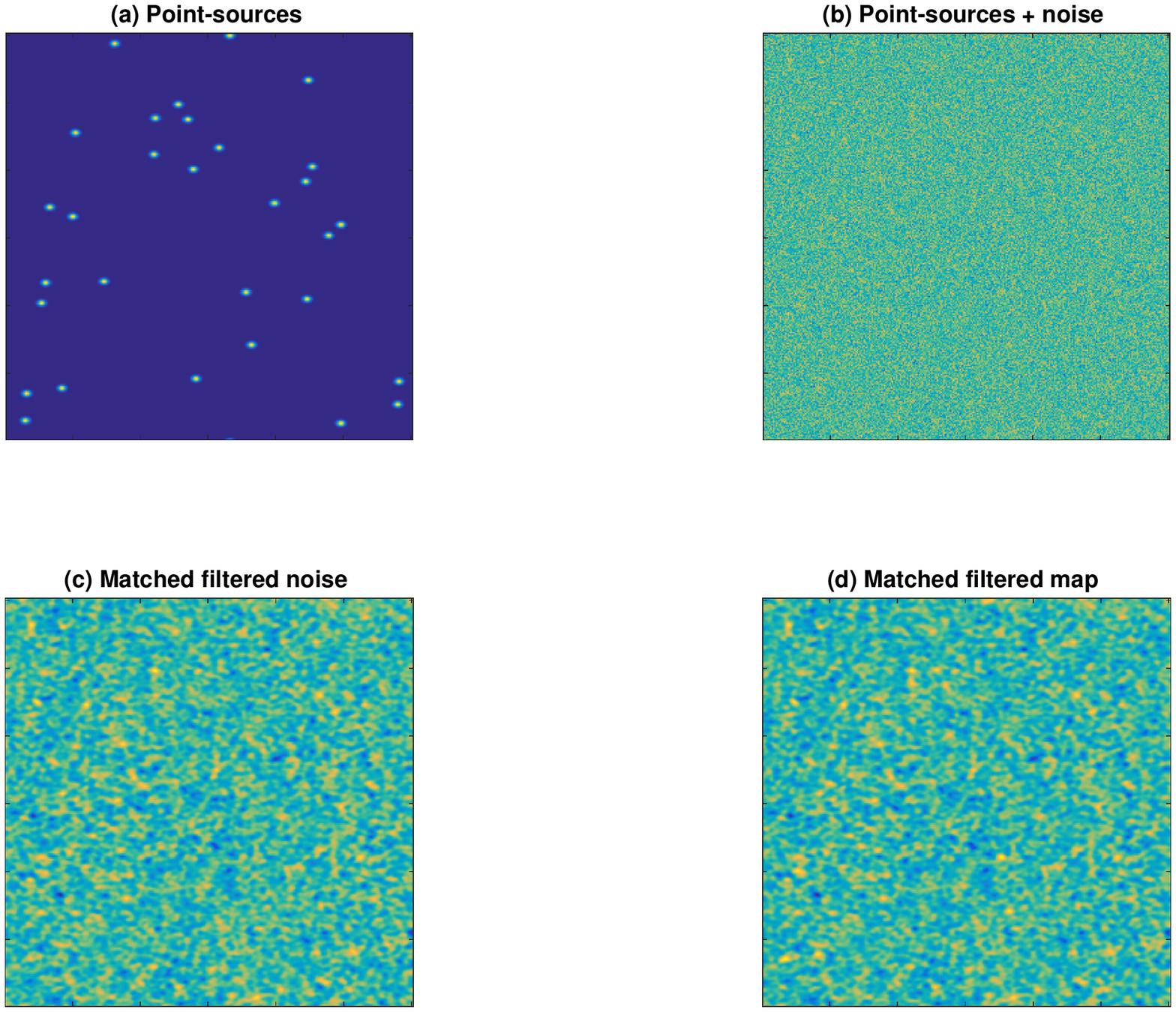}}
        \caption{(a) Simulated 30 point sources on a map of $300 \times 300$ pixels; (b) same as in panel (a) with the addition of a Gaussian white-type noise. The amplitude of the sources is well below the noise level; (c) zero-mean unit-variance 
version of the noise map after the application of the MF; (d) zero-mean unit-variance version of the map obtained after the application of the MF to the map in panel (b).}
        \label{fig03}
    \end{figure}
\end{landscape}

To illustrate the usefulness of the proposed method, this is applied to a simulated $300 \times 300$ pixels map where $30$ point sources, with a uniform random spatial distribution and Gaussian profile with standard deviation along the horizontal and vertical directions set to $1.3$ and $1.8$ pixels, respectively, (see Fig.~\ref{fig03}(a)), are
embedded in a Gaussian white noise. All the point sources have the same amplitude set to the standard deviation of the noise. This experiment simulates a very difficult situation where the amplitude of the point sources is
well below the level of the noise. Indeed, in Fig.~\ref{fig03}(b), which shows the observed map, the sources are not even visible.  In situations like this a matched filtering operation is unavoidable.
As seen above, the MF represents an optimal solution. However,  a comparison of Figs.~\ref{fig03}(c) and ~\ref{fig03}(d), which show
the  zero-mean unit-variance matched filtered versions of the noise component and observed map,  respectively, indicates that also after the MF operation the detection of the point sources is still a problematic issue since no strong peaks are evident.
Moreover, as discussed in Sect.~\ref{sec:morph}, after the matched filtering, the blob-shaped structures due to the noise have a shape similar to that of the filtered point sources, i.e.,
the morphological analysis cannot be used to detect the searched point sources.

Because of the asymmetric shape of the point sources,
the noise background is non-isotropic but simply homogeneous. This does not represent a problem given that the corresponding  two-point autocorrelation function is a two-dimensional Gaussian function, i.e., of the type $\rho(r)= f(\rb^T \Ab \rb)$
as in Sect.~\ref{sec:twodimensional}. A situation like this one reflects the results found in ALMA observations (see below). 

\begin{landscape}
   \begin{figure}
        \resizebox{\hsize}{!}{\includegraphics{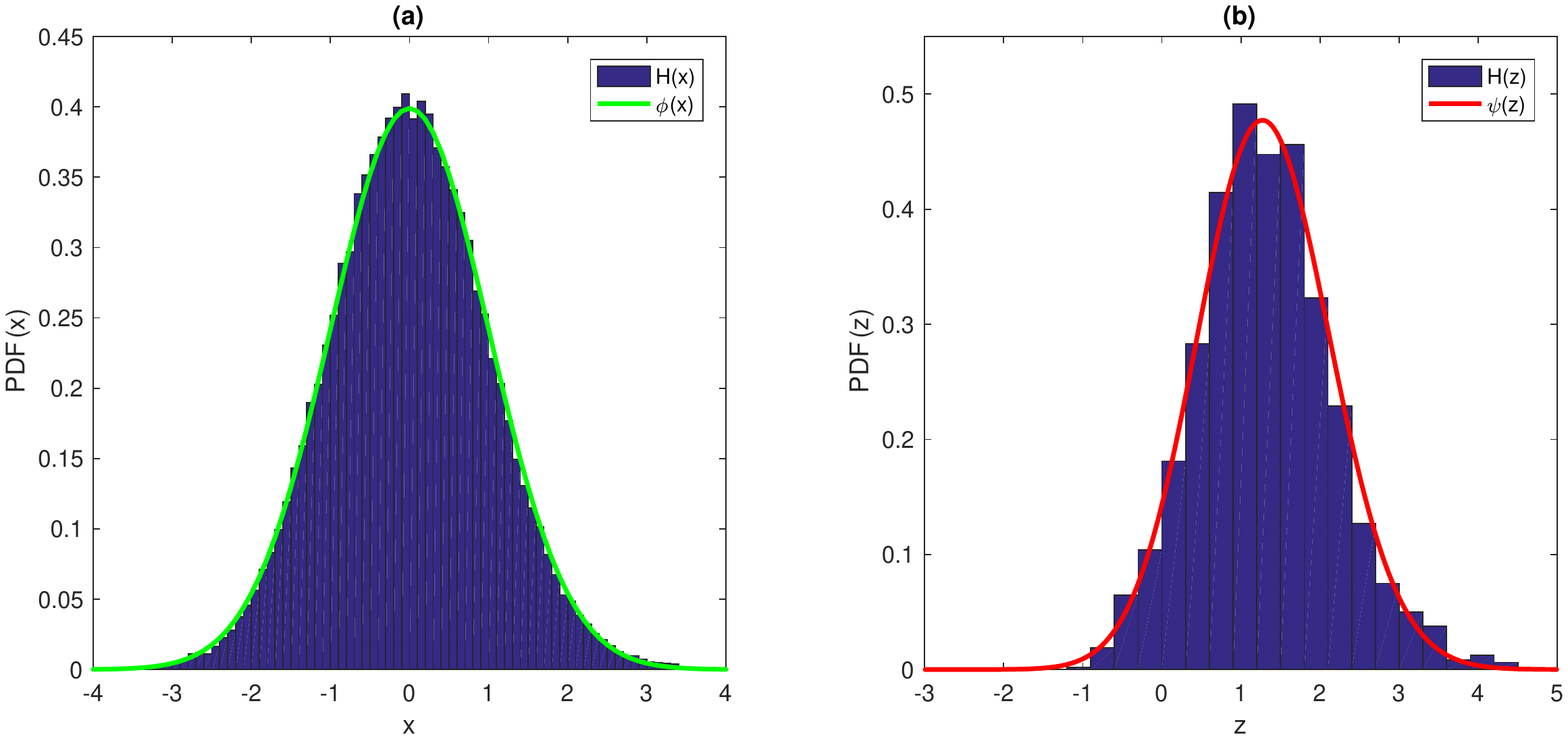}}
        \caption{(a) Histogram $H(x)$ of the pixel values of the map in Fig.~\ref{fig03}(d). For reference, the standard Gaussian PDF $\phi(x)$ is also plotted; (b) histogram $H(z)$ of the values of the peaks in the same map
vs. the corresponding PDF $\psi(z)$ is shown. There is good agreement between $\phi(x)$ and $\psi(x)$ with the corresponding histograms.} 
        \label{fig04}
    \end{figure}
\end{landscape}

\begin{landscape}
   \begin{figure}
        \resizebox{\hsize}{!}{\includegraphics{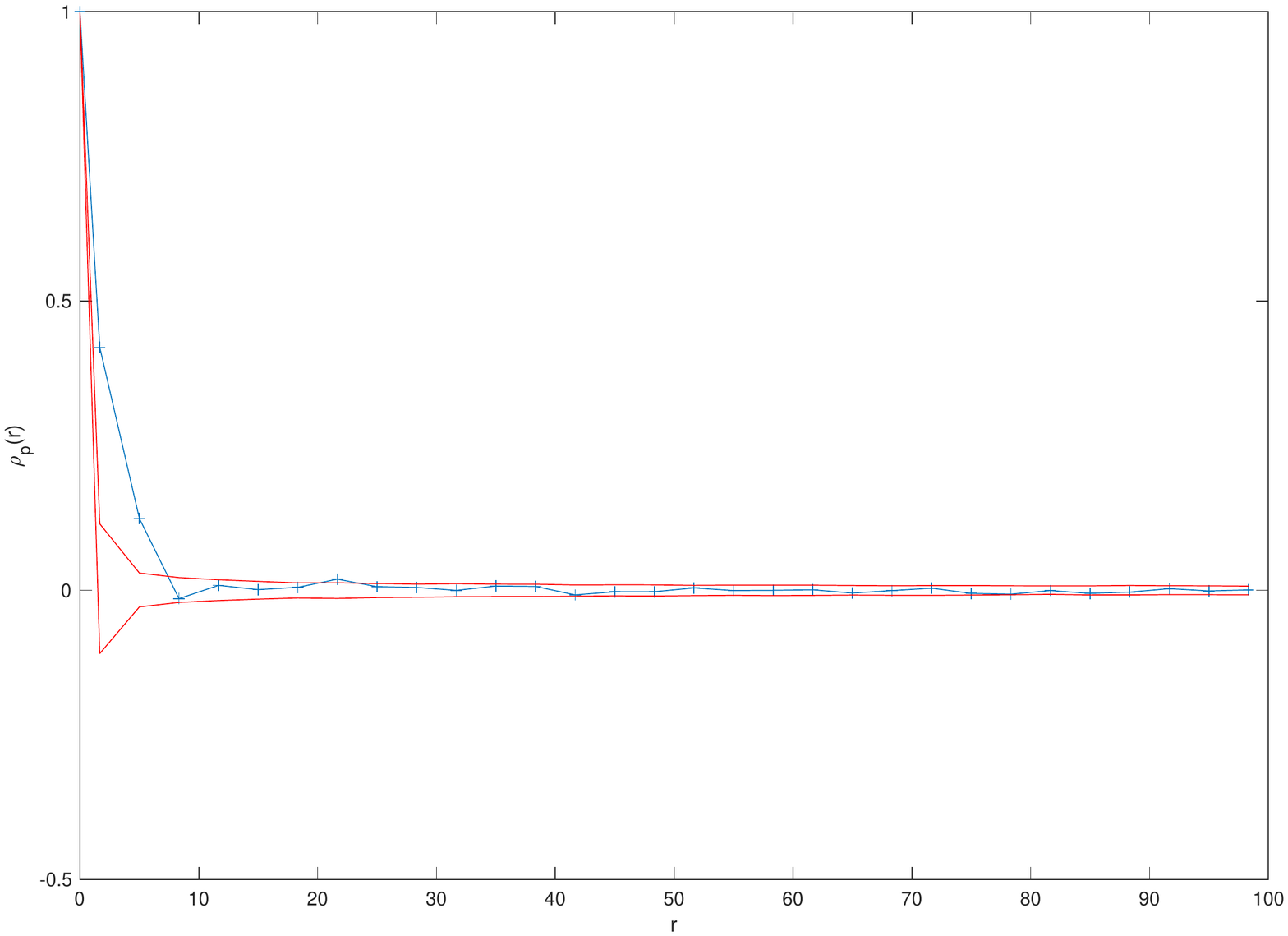}}
        \caption{Two-point correlation function $\rho_p(r)$ for the peaks in Fig.~\ref{fig03}(d). The inter-point distance $r$ is in pixels units. The two red lines define the $95\%$ confidence band. They were obtained by means of a bootstrap method based on the $95\%$ percentile envelopes of the two-point correlation functions obtained from $1000$ resampled sets of peaks with the same spatial coordinates as in the original signal but whose values were randomly permuted.}
        \label{fig05}
    \end{figure}
\end{landscape}

\begin{figure*}
        \resizebox{\hsize}{!}{\includegraphics{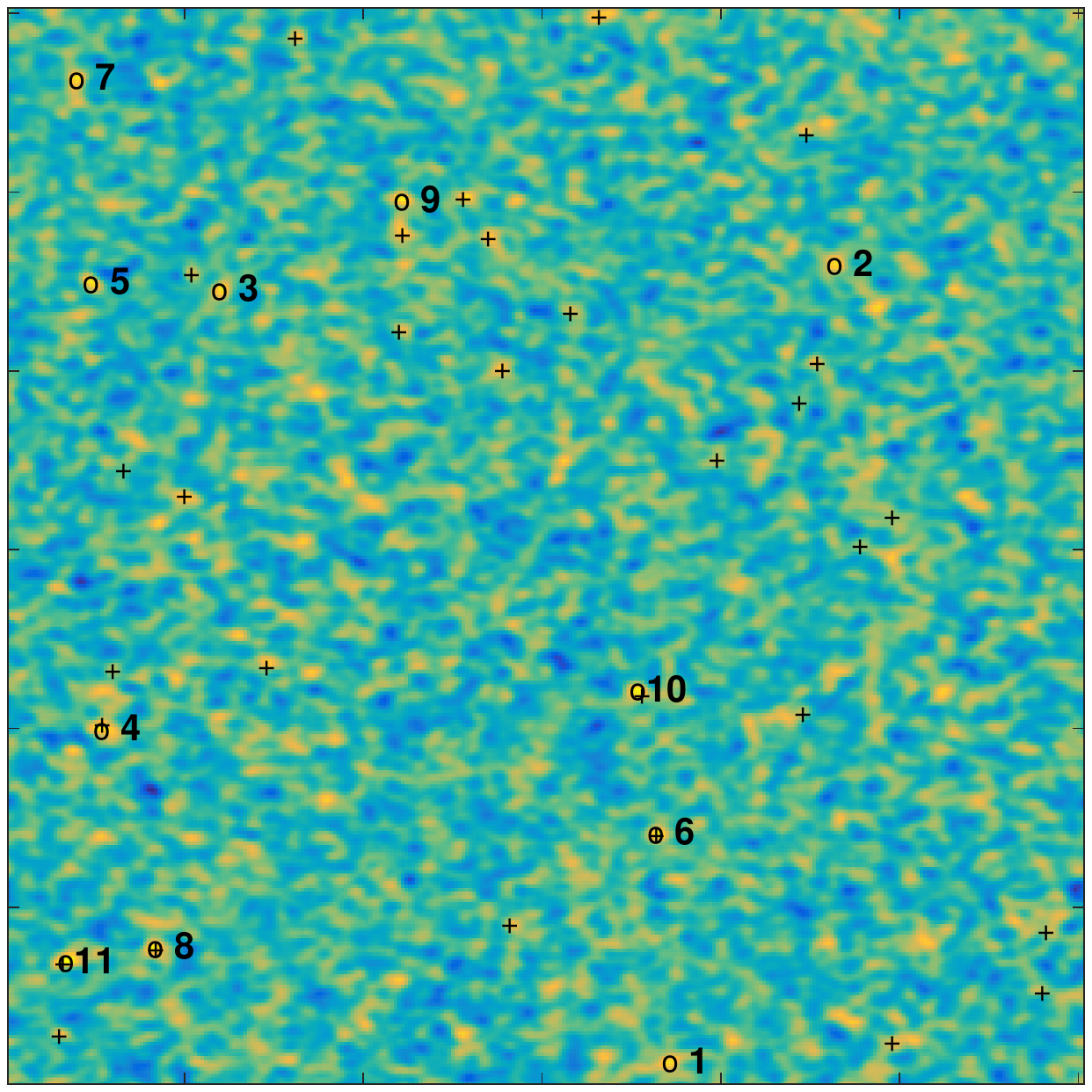}}
        \caption{Detection results for the peaks in Fig.~\ref{fig03}(d). The white circles highlight the peaks with a PFA smaller than $2.55 \times 10^{-3}$. Each of such peaks is
indexed with an increasing number according to its amplitude. The black crosses highlight the true position of the  point sources.}
        \label{fig06}
    \end{figure*}

The histogram of the pixel values of the matched filtered map
in Fig.~\ref{fig04}(a) is clearly compatible with a Gaussian PDF $\phi(x)$. The number of identified peaks is $1601$ and the estimated $\hat{\kappa}$ is about $1$.  Figure~\ref{fig04}(b) shows that the corresponding 
PDF $\psi(z)$ is in good agreement with the histogram $H(z)$ of the peak values.  The iid condition for the peaks, which is necessary for the computation of the PSFA, is demonstrated in Fig.~\ref{fig05},
which shows the two-point correlation function of the peaks. As a preliminary detection threshold, a value
of $u=3.72$ was chosen. Eleven peaks exceed it.  In Figs.~\ref{fig06} they are  highlighted and indexed with an increasing number according to the amplitude. Among these, only five ( \#4, \#6, \#8, \#10, and \#11)
correspond to a point source. In a situation like this one, an accurate quantification of the detection reliability is critical.

\begin{landscape}
   \begin{figure}
        \resizebox{\hsize}{!}{\includegraphics{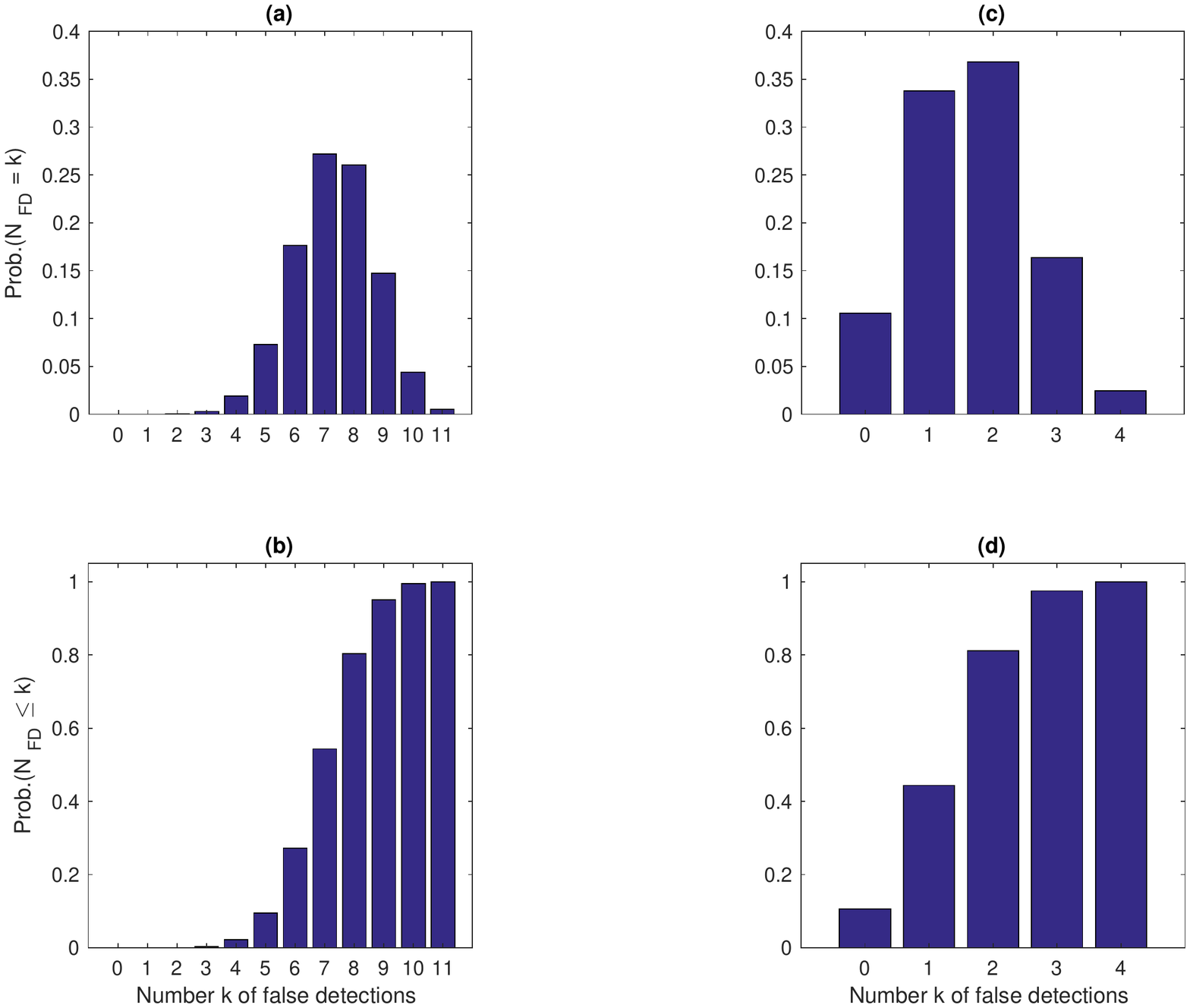}}
        \caption{(a) Probability distribution function (PDF)  of the number $N_{\rm FD}$ of false detections among the peaks \#1 - \#11 in Fig~\ref{fig06}; (b) cumulative distribution function (CDF) is shown corresponding to the PDF in the
previous panel; (c) PDF is shown of the number $N_{\rm FD}$ of false detection among the peaks \#8 - \#11 in Fig~\ref{fig06}; CDF is shown corresponding to the PDF in the previous panel.}
        \label{fig07}
    \end{figure}
\end{landscape}

Since for the PFA~\eqref{eq:fd1} it is $\Phi_c(3.72)\approx 10^{-4}$, according to the standard procedure all the peaks should be considered true detections with a high
confidence level. To a minor extent, the same also holds for the the PFA~\eqref{eq:corra} since $\Psi_c(3.72)\approx 2.55 \times 10^{-3}$.  On the other hand, with $u=5$, an often used threshold, no peak should have been selected.
This supports the unreliability of a detection based only on the threshold $u$.
The situation with SPFA is different. From Fig.~\ref{fig07}(a), which shows the Poisson-binomial distribution corresponding to the SPFAs of the selected peaks, it appears that the most probable number $N_{\rm FD}$ of false detections is
$N_{\rm FD}=7$. Moreover, Fig.~\ref{fig07}(b) shows that the probability that $N_{\rm FD} \leq 7$ is about $0.54$. This result is in good agreement with the true $N_{\rm FD} =6$. 
If only the four highest peaks \#8 - \#11 are considered, with values of the SPFA equal to
$0.60$, $0.42$, $0.40$, and $0.24$, the most probable $N_{\rm FD}$ is $2$ and the probability that $N_{\rm FD} \leq 2$ is about $0.81$ (see Figs.~ \ref{fig07}(c)-(d)).
Among the four peaks, only the peak \#9 corresponds to a false detection, again in good agreement with the expected number. 

It is useful to compare the proposed method with the popular approach of the least-squares fit, on a small submap centered on a given peak, of a model constituted by the template $\gb$ superimposed on a two-dimensional low-degree  polynomial . According to this approach, the reliability of the detection is tested by comparing the estimated amplitude with a multiple of the standard deviation $\sigma_r$ of the residuals. 
When applied to the $11$ peaks selected above, assuming a two-dimensional polynomial of first degree and using a submap of size $11 \times 15$ pixels, 
in all of the cases the estimated amplitude is greater than the threshold set to $5 \sigma_r$. As a consequence, all of the peaks should be considered true detections with an high degree of reliability. Two more true point sources have been detected,
but at the cost of six false detections. 

\begin{figure*}
        \resizebox{\hsize}{!}{\includegraphics{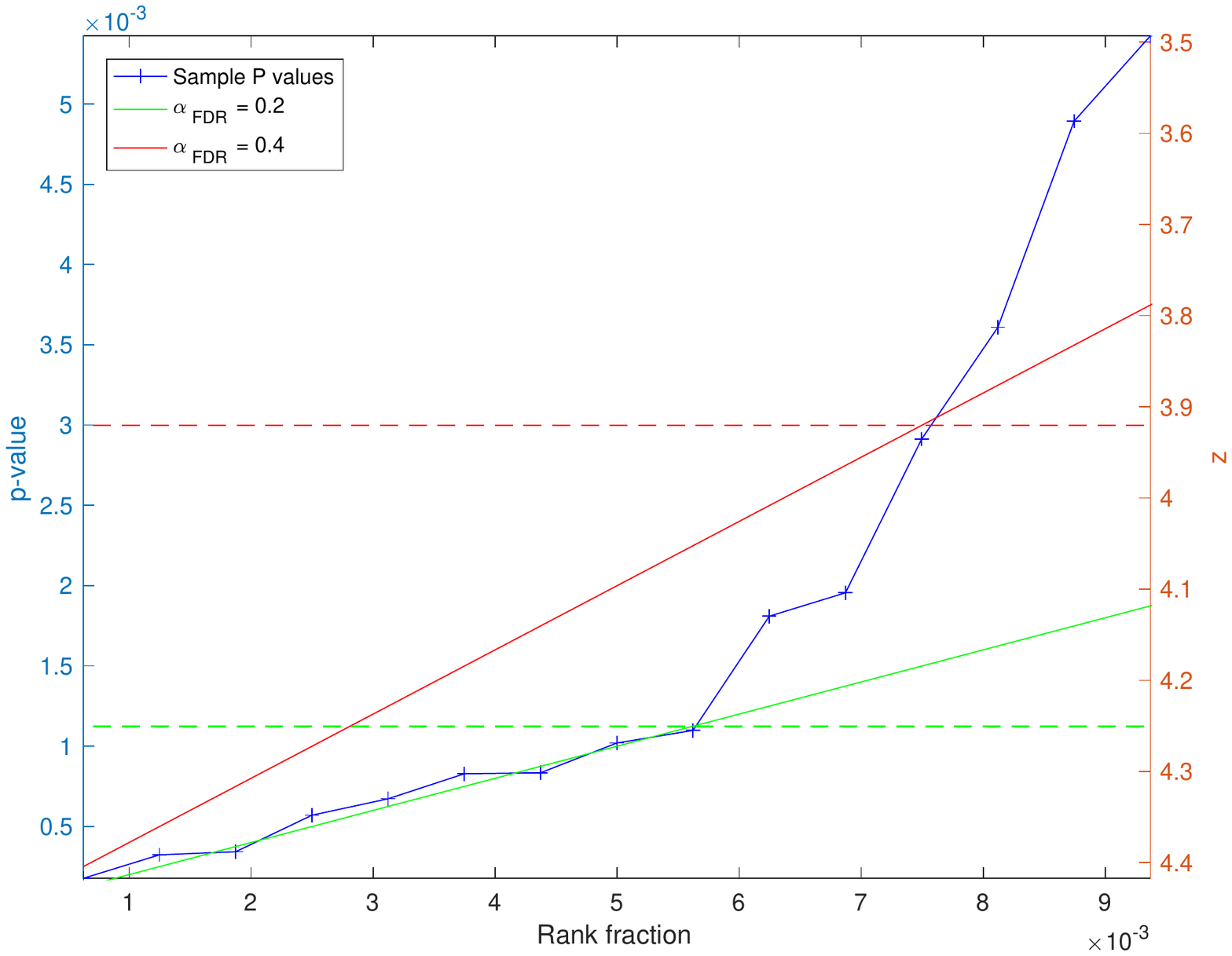}}
        \caption{Rank fraction vs. p-values (left axis) and peak amplitudes (right axis) of the peaks in Fig.~\ref{fig03}(d) (see text). Here the p-values correspond to the PFA of the peaks. The dashed lines provide the detection threshold for 
the corresponding  false discovery rate $\alpha_{\rm FDR}$. Any cross below a detection threshold corresponds to a detection. The detection thresholds are determined by the greatest p-value below the corresponding continuous line 
\citep[for details, see ][]{mil01}.}
        \label{fig08}
    \end{figure*}

Another useful comparison is with the FDR technique mentioned in Sect.~\ref{sec:spfa}. With this approach it is possible to control the expected number of detections falsely declared significant as a proportion $\alpha_{\rm FDR}$ of the number of all tests declared significant. Figure~\ref{fig08} shows that with $\alpha_{\rm FDR}=0.2$ and $0.4$ the number of detected point sources is $9$ and $12$, respectively.
However, in the first case the number of false detection is $4$, whereas it is $7$ in the second case. In both cases more true point sources have been detected
than the methodology based on the PDF $g(z_{\max})$,  but at the cost of a larger fraction of false detections. Moreover, also here, no quantification is possible for the reliability of each specific detection.
Finally, it is worth noticing that the actual fraction of false detections is $0.44$ and $0.58$, which is well above the corresponding expected value $\alpha_{\rm FDR}$.

The conclusion is that only the SPFA is able to really quantify the true reliability of a given detection and should be considered before follow-up observations to identify the source itself.

\section{Application to real ALMA maps} \label{sec:real}

\begin{figure*}
        \resizebox{\hsize}{!}{\includegraphics{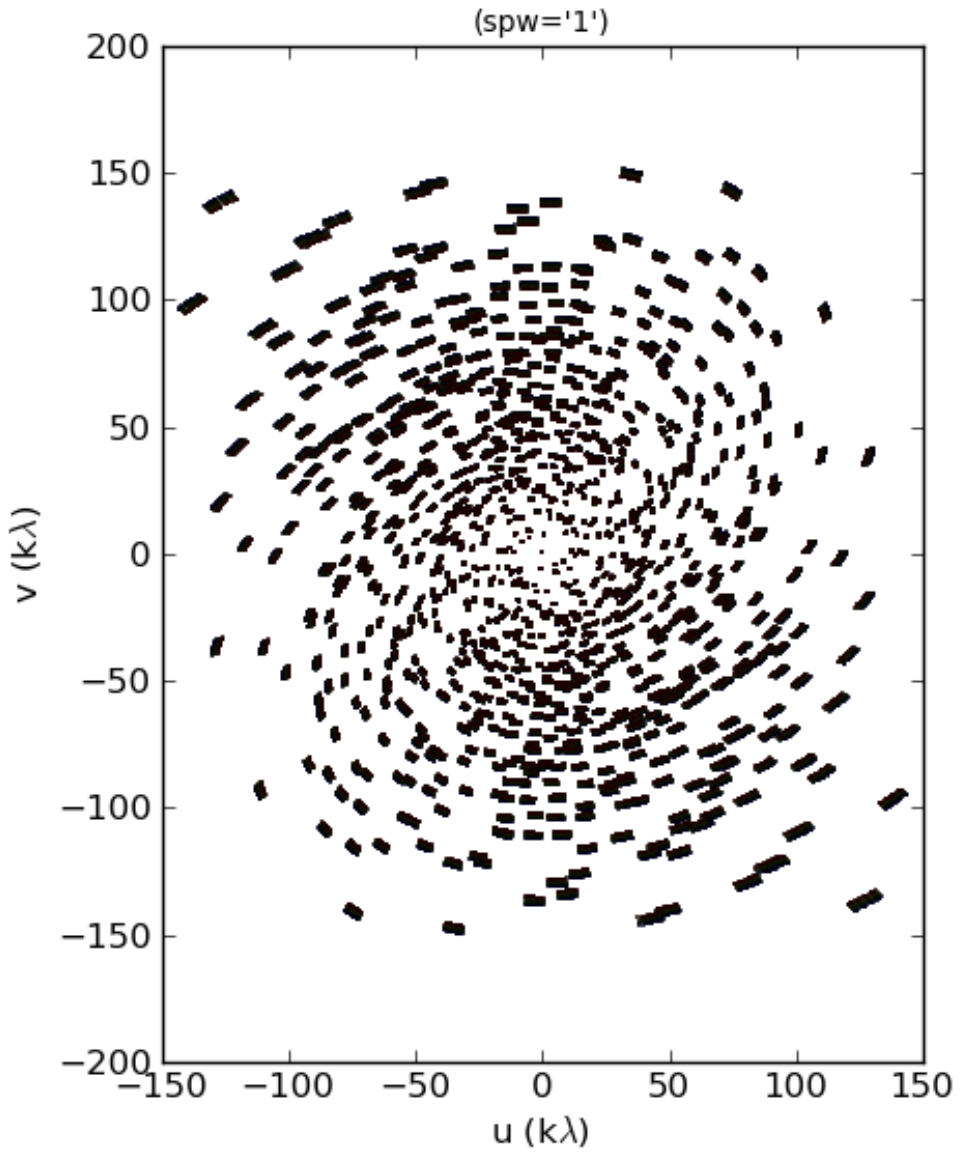}}
        \caption{uv-plane coverage of the ALMA band 3 observations of the radio source field PKS0745-19.}
        \label{fig09}
    \end{figure*}

We apply the aforementioned procedure to two interferometric maps, obtained at two frequencies with ALMA, with the aim to detect the faint point sources in the field of the radio source \object{PKS0745-191}. These maps are characterized
by an excellent {\it uv}-plane coverage (see Fig.~\ref{fig09}). For this reason, the noise is expected to have a uniform spatial distribution over the entire area of interest that is the basic condition for the proposed method to work.
The first map, (hereafter M1) was obtained at a frequency of $100 {\rm GHz}$ ($3 {\rm mm}$, ALMA band 3), whereas the second map (hereafter M2) was obtained at a frequency of 
$340 {\rm GHz}$ ($0.87 {\rm mm}$, ALMA band 7).
Both maps, with a size of $256 \times 256$ pixels, are centered on the bright cluster galaxy (BCG) of PKS0745-19 (RA 07:47:31.3, Dec -19:17:39.94, J2000), which was the target of the observations we used (ALMA ID = 2012.1.00837.S, PI R. McNamara). The data reduction was carried out with CASA version 4.2.2 and the ALMA reduction scripts \citep{mcm07}.
To produce the continuum maps, we selected and averaged only the channels free of the CO line emission, which was the original observation target.
The images were reconstructed via the CASA task \textit{clean} with Briggs weighting and a robust parameter of 1.5 in band 3 and 2 in band 7.
The beam size is 1.9” $\times$ 1.4” in band 3, and 0.27” $\times$ 0.19” in band 7.

\begin{landscape}
   \begin{figure}
        \resizebox{\hsize}{!}{\includegraphics{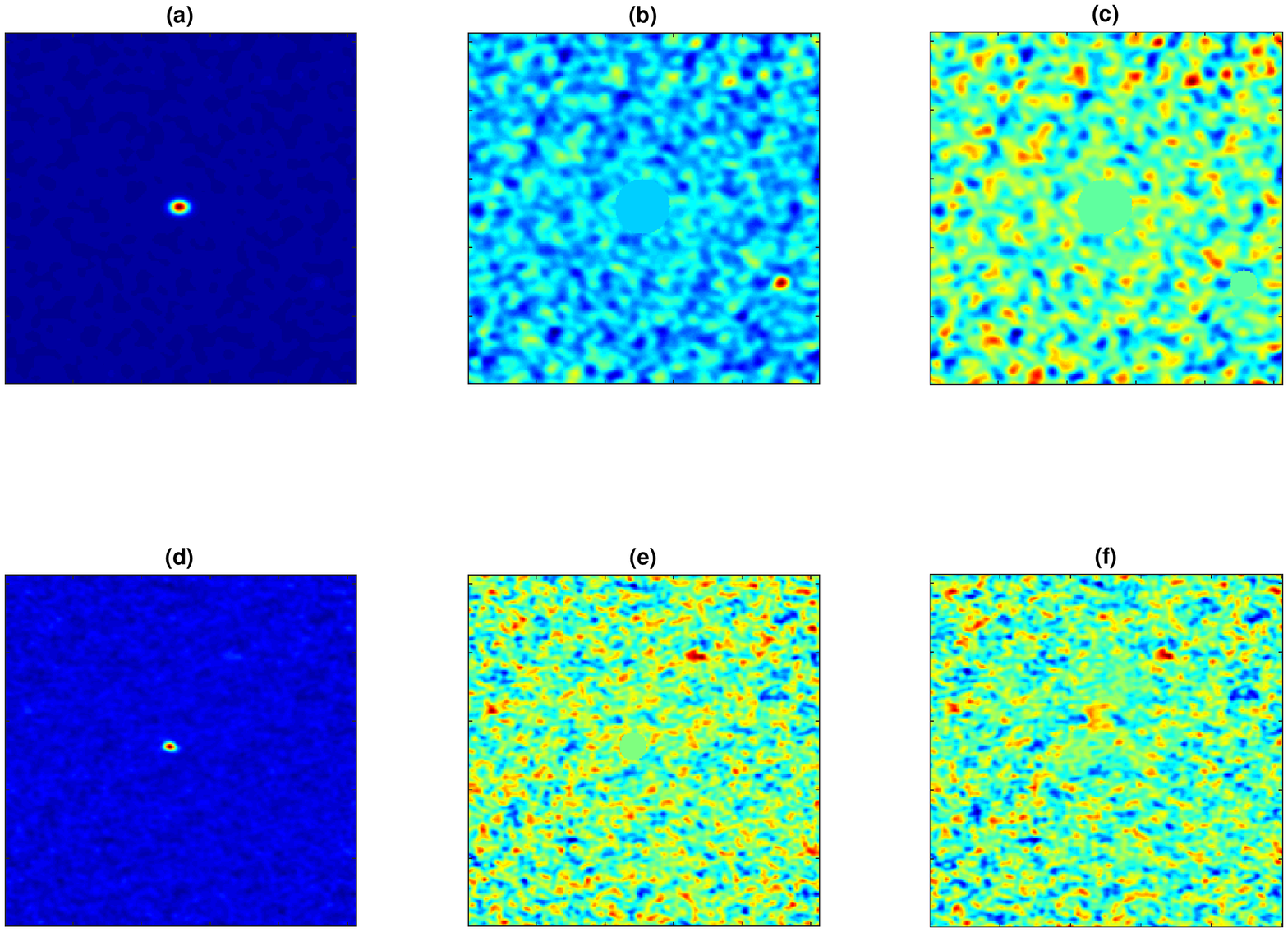}}
        \caption{(a) Original ALMA map M1 is shown; (b) map M1 with the brightest sources masked is shown; (c) map M1 with both brightest sources masked is shown; (d) original ALMA map M2 is shown; (e) map M2 with the brightest sources masked
is shown; (f) map M3, obtained by composing M1 and M2 with the Feather algorithm implemented in CASA, is shown.}
        \label{fig10}
    \end{figure}
\end{landscape}

In Figs.~\ref{fig10}(a) and (d) a bright source is visible in the central position of both maps, which corresponds to the BCG. Figures~\ref{fig10}(e) and (b) show that, when this is masked in M2 no additional sources become visible whereas a bright source appears in M1. This additional source has been identified as an IR source already known in the cluster,  i.e., \object{2MASX J07473002-1917503} (RA 17:47:30.130, Dec -19:17:50.60, J2000). If this is also masked in M1 no additional sources are obviously visible in Fig.~\ref{fig10}(c). Since we are interested in the detection of point sources with amplitudes comparable to the noise level, the pixels in the masked areas are not used in the analysis.

\begin{landscape}
   \begin{figure}
        \resizebox{\hsize}{!}{\includegraphics{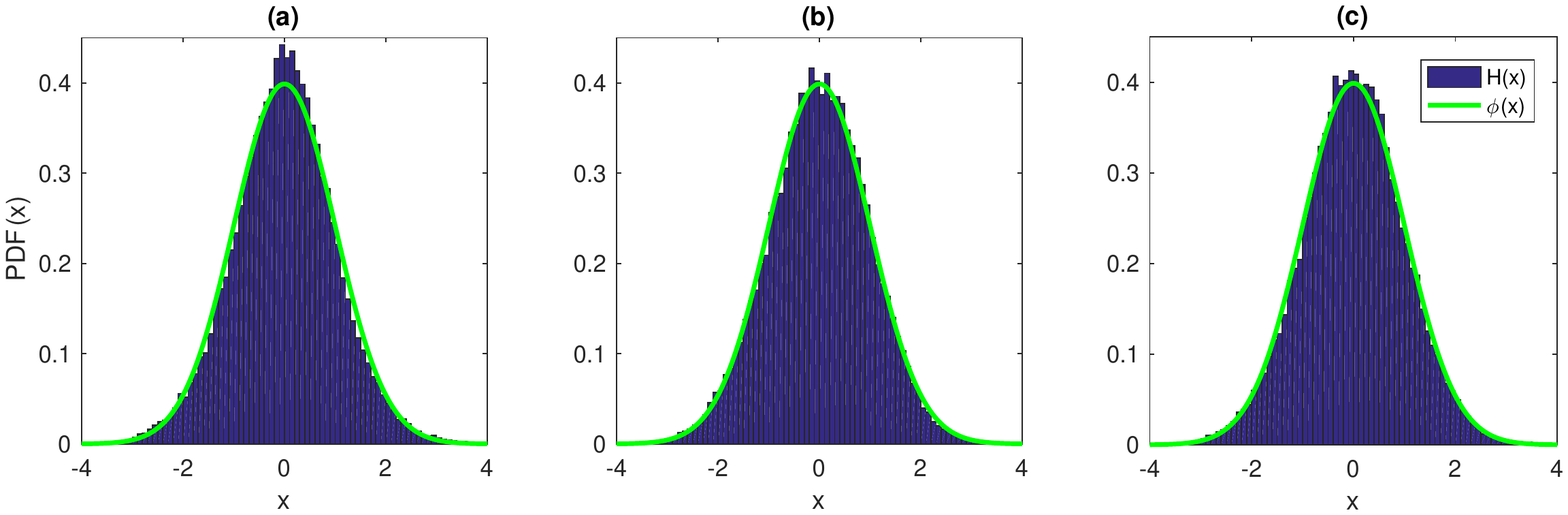}}
        \caption{(a) Histogram $H(x)$ of the pixel values of the map M1, (b) M2 and (c) M3. For reference, the standard Gaussian PDF $\phi(x)$ is also plotted.}
        \label{fig11}
    \end{figure}
\end{landscape}

\begin{landscape}
   \begin{figure}
        \resizebox{\hsize}{!}{\includegraphics{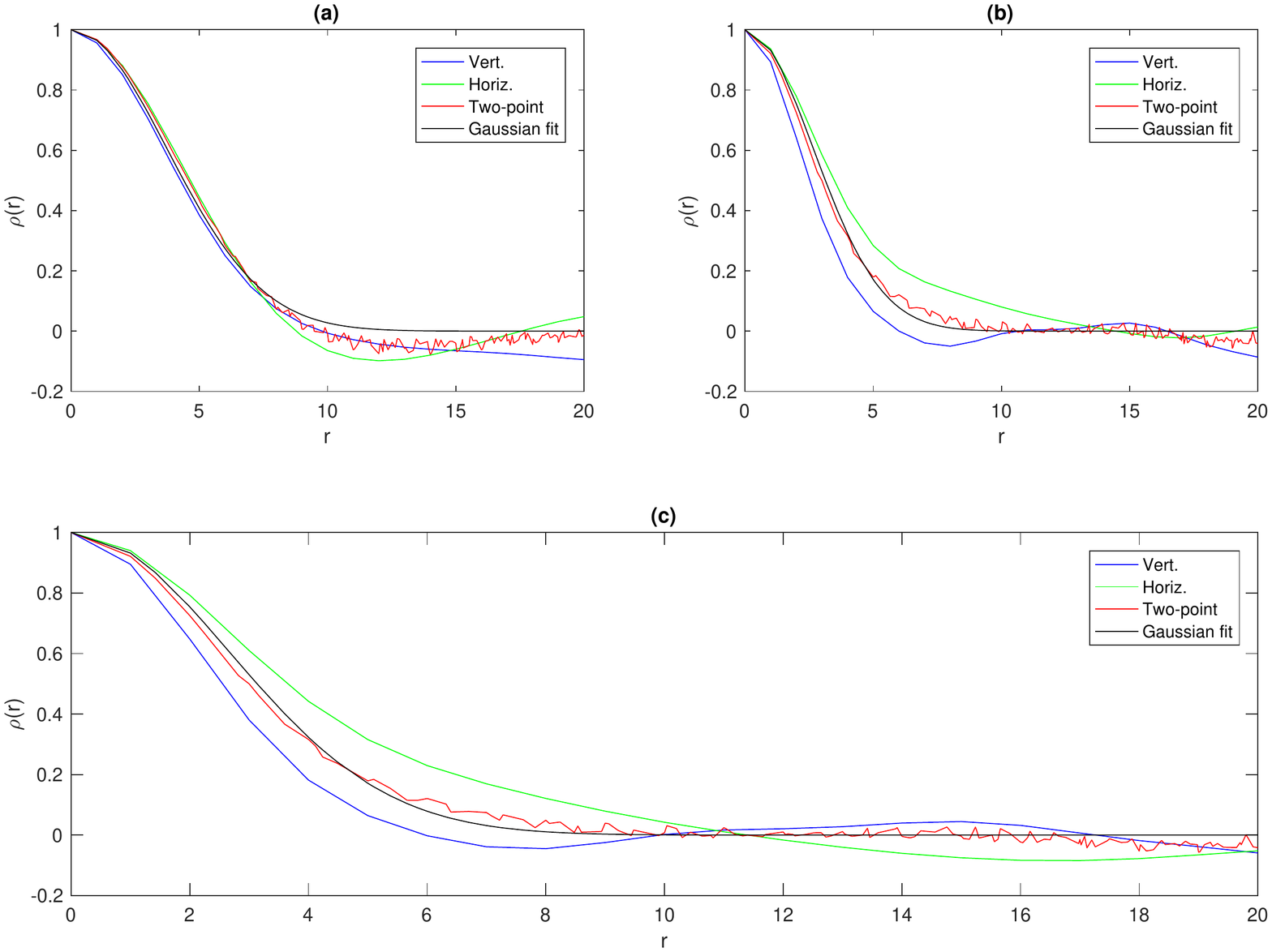}}
        \caption{(a) Autocorrelation function along the vertical and horizontal directions, two-point correlation function, and the corresponding least-squares fit with a Gaussian function of the map M1 are shown. The inter-pixel distance $r$ is in pixels units; 
        (b)-(c) the same for the maps M2 and M3.}
        \label{fig12}
    \end{figure}
\end{landscape}

Most of the structures visible in Figs.~\ref{fig10}(c) and (e) are certainly not due to physical emission but to the noise. In Figs.~\ref{fig11}(a)-(b), the histograms $H(x)$ of the pixel values of these maps,
standardized to zero mean and unit variance, indicate that the noise is Gaussian but, as shown by the two-point autocorrelation functions in Figs.~\ref{fig12}(a)-(b), not white. 

As a first step, the two maps were independently analyzed. As already underlined by \citet{vio16}, the use of the MF with these kinds of ALMA maps is unfeasible because the MF filters out the Fourier frequencies where the noise is predominant, preserving those where the searched signals give a greater contribution.
However, the unresolved point sources and the blob-shaped structures due to the noise have similar appearances and the process of filtering cannot work.
In this case $\Tc(\xb, i_0) = \xb$ and the detection test becomes a thresholding test according to which a peak in the map can be attributed to a point source 
if it exceeds a given threshold. To apply the procedure introduced in Sect.~\ref{sec:twodimensional} it is necessary to test the isotropy of the noise field.
From Figs.~\ref{fig12}(a)-(b) this condition appears to be approximately satisfied. The small differences between the autocorrelation functions along the vertical and horizontal directions is due to the elliptical shape of the ALMA beam.

\begin{landscape}
   \begin{figure}
        \resizebox{\hsize}{!}{\includegraphics{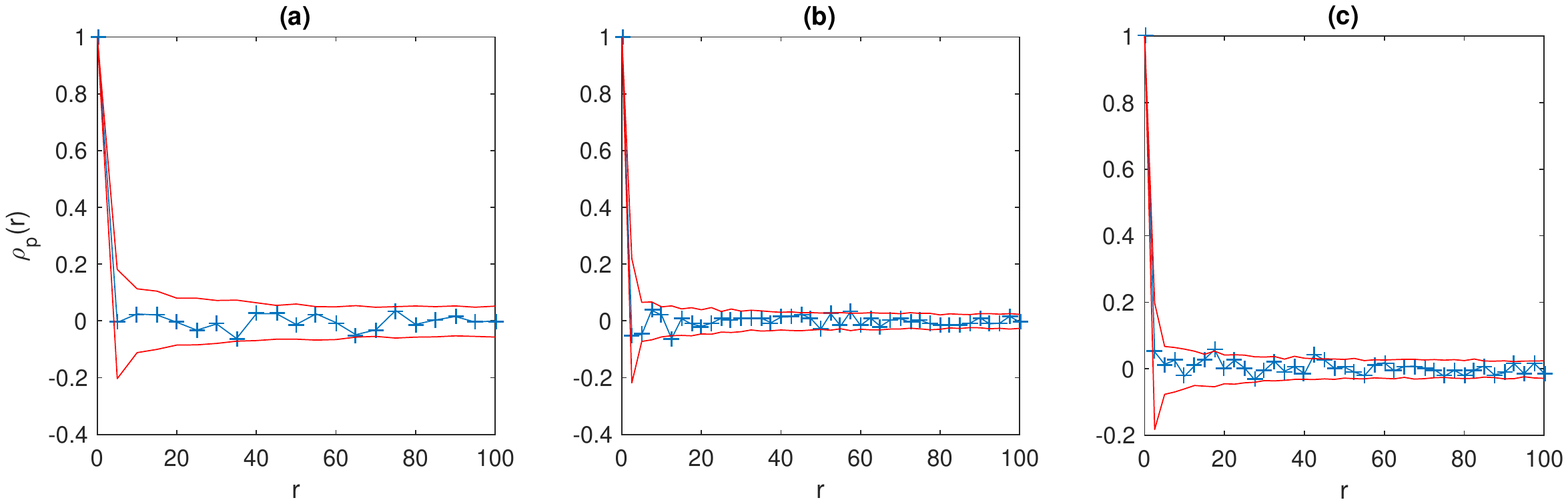}}
        \caption{(a) Two-point correlation function $\rho_p(r)$ for the peaks in the map M1 is shown. The inter-point distance $r$ is in pixels units. (b)-(c) The same for the maps M2 and M3. The two red lines define the $95\%$ confidence band. These were obtained by means of a bootstrap method based on the $95\%$ percentile envelopes of the two-point correlation functions obtained from $1000$ resampled sets of peaks with the same spatial coordinates as in the original signal but whose values were randomly permuted.}
        \label{fig13}
    \end{figure}
\end{landscape}

\begin{landscape}
   \begin{figure}
    \resizebox{\hsize}{!}{\includegraphics{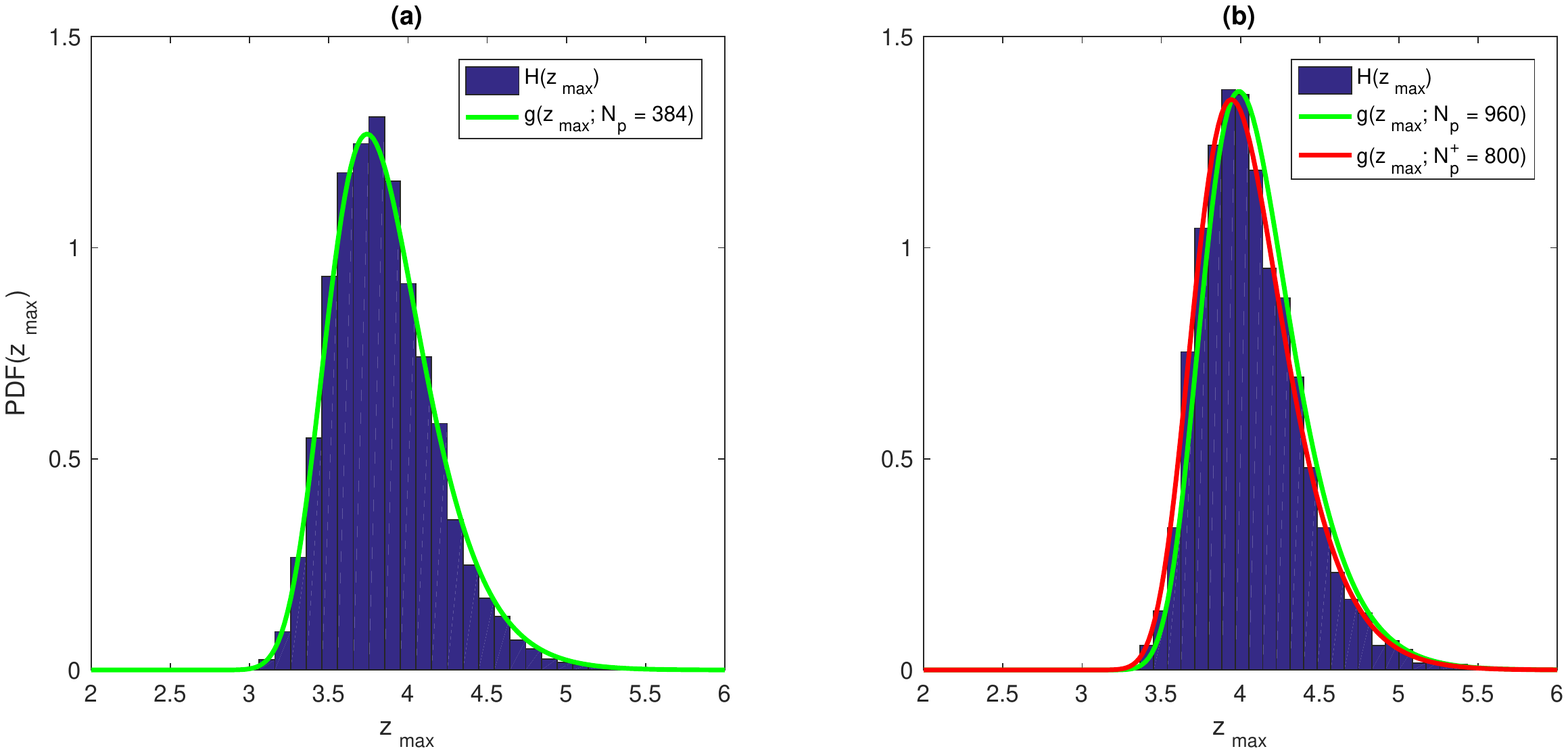}}
        \caption{(a)-(b) Histograms vs. the PDFs $g(z_{\rm max})$ of the largest peak value from $5000$ Gaussian random fields obtained by filtering a $256 x 256$ discrete white noise map by means
        of a Gaussian filter with dispersion of $3.7$ and $2.7$ pixels in such a way as to reproduce approximately the noise in M1 and M2, respectively. Since each simulated map is characterized by a different number of peaks, the corresponding PDFs
       are slightly different from one another. For this reason, the displayed  $g(z_{\rm max})$ plotted in green corresponds to the mean number of peaks which are $384$ and $960$, respectively. In panel (b) the resulting
       PDF is in reasonably good agreement with the corresponding histogram, but a better result is obtainable if an effective number $N_p^+ = 800$ is used. To stress the fact that,
       although $N_p^+$ is about $17\%$  smaller than $N_p$, the resulting PDF is only slightly different. All this is in accordance with the arguments in Sect.~\ref{sec:spfa}.}
        \label{fig14}
    \end{figure}
\end{landscape}

We identified $328$ peaks in M1 and $948$ in M2. The iid condition for the peaks is supported by the two-point correlation functions $\rho_p(r)$ in  Figs.~\ref{fig13}(a) and (b), which are almost completely contained in their $95\%$ confidence band. 
We obtained the confidence band by means of a bootstrap method based on the $95\%$ percentile envelopes of the two-point correlation functions computed from $1000$ resampled sets of peaks with the same spatial coordinates as in the original map but whose values have been randomly permuted.
The reliability of the iid condition is confirmed by Fig.~\ref{fig14}, which shows the PDF of the largest peak value from a set of $256 x 256$ pixels Gaussian random fields obtained by filtering $5000$ discrete white noise maps by means of a Gaussian filter with dispersion of $3.7$ and $2.7$ pixels in such a way to approximately reproduce the noise in M1 and M2, respectively.

\begin{landscape}
   \begin{figure}
        \resizebox{\hsize}{!}{\includegraphics{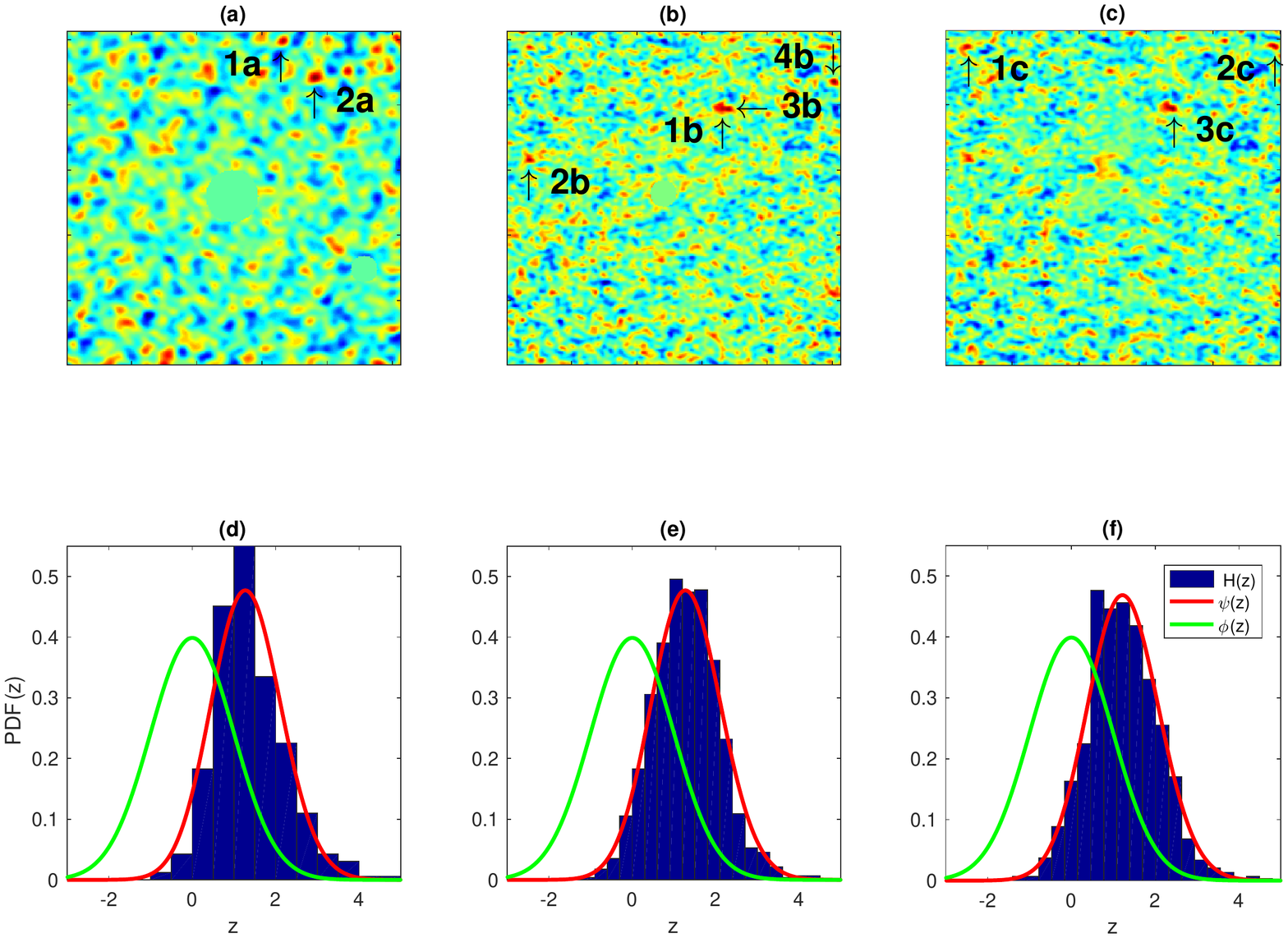}}
        \caption{(a)-(c) Detected sources with a probability of false detection smaller than $10^{-3}$ on the maps M1, M2 and M3, respectively; (e)-(f) corresponding histograms $H(z)$ and PDFs $\psi(z)$ of the peaks are shown. 
	For reference, the Gaussian PDF $\phi(z)$ is also plotted.}
        \label{fig15}
    \end{figure}
\end{landscape}

The maximum likelihood estimate~\eqref{eq:ml} provides $\hat{\kappa}\approx 1$ for both M1 and M2, which is a value that is typical of Gaussian two-point autocorrelations functions. This is an expected result for
ALMA given that, independently of the frequency, its PSF can be well approximated by bivariate Gaussian functions. The least-squares fit of the two-point correlation function $\rho(r)$ 
in Figs.~\ref{fig12}(a)-(b) with a Gaussian function supports this circumstance. As seen at the end of Sect.~\ref{sec:twodimensional}, this fact makes even more irrelevant the small anisotropy of noise background observed above.
The  PDFs $\psi(z)$ are shown in Figs.~\ref{fig15}(c) and (d). The agreement with the respective histograms $H(z)$  is good. A  threshold $u \approx 3.98$, corresponding to a PFA~\eqref{eq:corra}  equal to $10^{-3}$, provides two candidate point sources in M1 and four in M2. All these detections are highlighted in Figs.~\ref{fig15}(a) and (b).
They are indexed with an increasing number according to the source amplitude.

\begin{landscape}
   \begin{figure}
        \resizebox{\hsize}{!}{\includegraphics{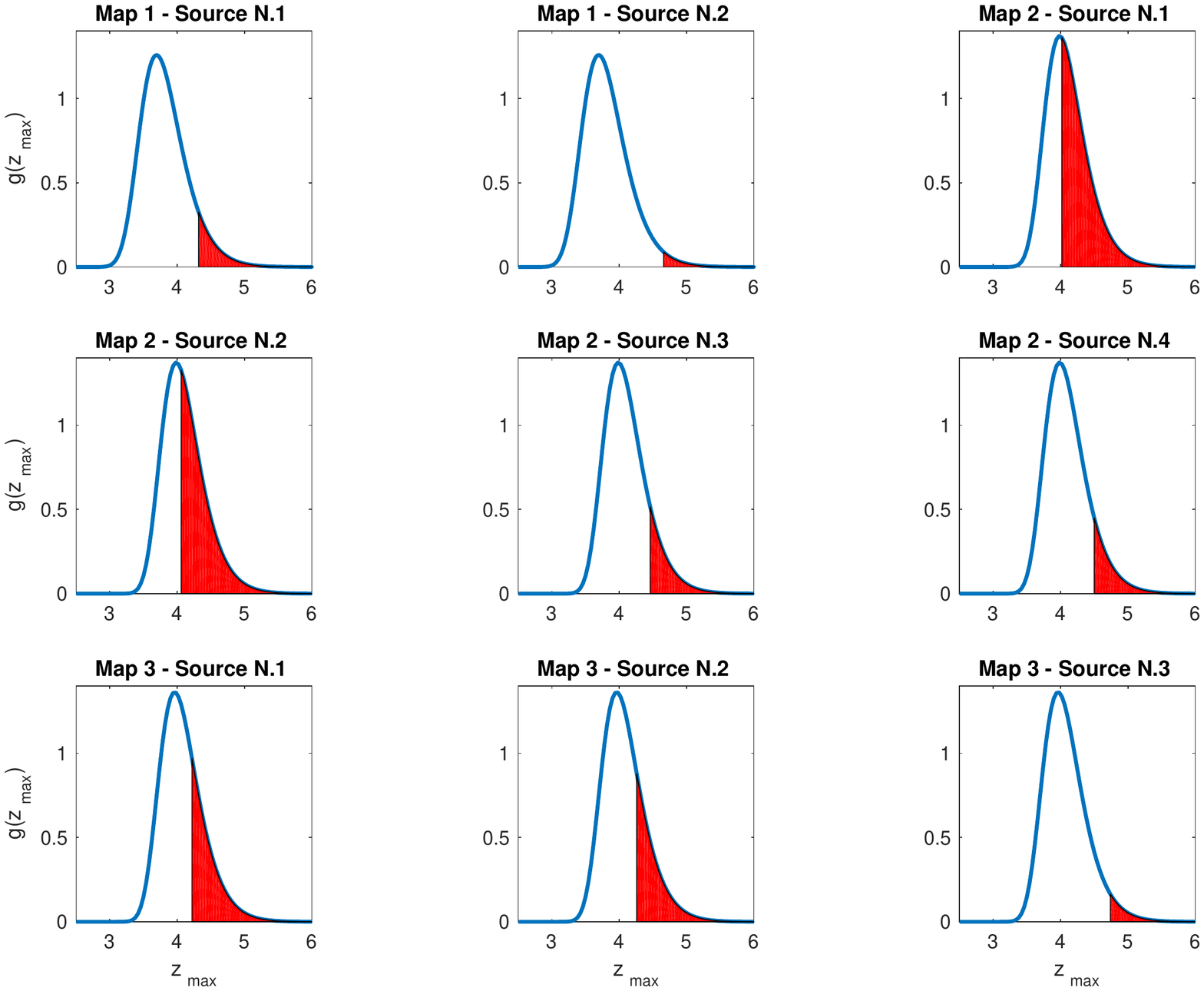}}
        \caption{Graphical representation (red-filled areas) of the SPFA for the detections with PFA $\le 10^{-3}$ in M1, M2, and M3.}
        \label{fig16}
    \end{figure}
\end{landscape}

Even though the PFA is identical for all the sources, their detection reliability is different. Indeed, in M1 the SPFA is $8.8 \times 10^{-2}$ and $2.0 \times 10^{-2}$ for
the sources 1a and 2a, whereas in M2 it is $5.7 \times 10^{-1}$, $5.0 \times 10^{-1}$, $1.4 \times 10^{-1}$, and $1.3 \times 10^{-1}$ for the sources 1b-4b, respectively (see also Fig.~\ref{fig16}).
The Poisson-binomial distribution corresponding to these SPFAs indicates that for M1 the probability of $N_{\rm FD}=0$ is about $0.89$, whereas for M2 the probability that $N_{\rm FD} \leq 1$
is $0.58$ and becomes $0.92$ for $N_{\rm FD} \leq 2$.

As a final comment, we note that if the standard PFA~\eqref{eq:fd1} based on the Gaussian PDF $\phi(z)$ had been used, the chosen detection threshold $u$ 
should have produced $104$ detections in M1 and $294$ in M2.

Source 1a in M1 was identified as an already known object \object{USNO B1.0 0707-10151219} (RA 07:47:30.817, Dec -19:17:18.48, J2000). On the map M2, the two peaks 1b and 3b look like a single extended source or two very close sources. However, the separation of the two peaks  is only $0.022'' \times 0.04''$, which is smaller than the beam size
and the two sources cannot be resolved. Therefore, these two peaks are probably due to a single extended source.  Table~\ref{list} summarizes our results.

\begin{table*}
\noindent \centering{}\protect\caption{List of detected sources}\label{list}
\begin{center}
\begin{tabular}{|c|c|c|c|c|}
\hline \hline
 & &  & &   \\
  Source ID & RA & Dec & Size (arcsec $\times$ arsec)& Reference  \\
  \hline
& &  & &   \\
  \textbf{Map M1/Band 3} & & & &  
\\
  \hline
  1a & 07:47:30.817 &  -19:17:18.48 & 1.76"$\times$1.57"  & USNO B1.0 0707-10151219\\
  \hline
  2a & 07:47:30.520 & -19:17:23.39 & 2.01"$\times$1.85" & unknown   \\
  \hline
  Bright source in Figure~\ref{fig10}(b) & 17:47:30.130 & -19:17:50.60 & 2.56"$\times$2.42" & 2MASX J07473002-1917503\\
  \hline
& &  & &   \\
  \textbf{Map M2/Band 7} & & & &
\\
  \hline
  1b &	07:47:31.206 & -19:17:37.38 & 0.32"$\times$0.26" & unknown  \\
  \hline
  2b & 07:47:31.642	& -19:17:38.85	& 0.52"$\times$0.40" & unknown \\
  \hline
  3b & 07:47:31.228 & -19:17:37.35	& 0.35"$\times$0.34" & unknown\\
  \hline
  1b+3b (as a single source) & 07:47:31.220 &	-19:17:37.35 & 0.44"$\times$0.32" & unknown \\
 \hline
  4b & 07:47:30.982 & -19:17:36.53 & 0.34"$\times$0.24" & unknown \\
\hline
\hline
& &  & &   \\
  \textbf{Map M3/Band 3 + Band 7} & & & &
\\
  \hline
\hline
1c	 & 07:47:31.618 & 	-19:17:35.464 &	0.39"$\times$0.18" & 	unknown \\
\hline
2c &	07:47:30.974 &	-19:17:35.464 & 0.41"$\times$0.24" &	unknown \\
\hline
3c &	07:47:31.192 &	-19:17:37.33 & 0.57"$\times$0.31" & same as 1b+3b in M2 \\

\hline 
\hline

\end{tabular}
\end{center}
\end{table*}

We computed the fluxes with the CASA statistics tool.
Since our sources are not resolved, to estimate the flux we  measured the integrated flux over the PSF, which is assumed to have a Gaussian shape. For each source, we assign a
root mean square (RMS) value computed over the pixels around the source, which is a rough indication of the noise level. Results are presented in Tab.~\ref{flux1}.
\begin{table*}
\noindent \centering{}\protect\caption{Source fluxes computed on the original images}\label{flux1}
\begin{center}
\begin{tabular}{|c|c|c|}
\hline \hline
& &    \\
 \multicolumn{3}{|c|}{\textbf{Map M1/Band 3}} \\
 \hline
& &   \\
Source ID & Integrated flux (mJy) & RMS (mJy) \\
\hline
BCG PKS0745-191 & 9.00 & 0.18 \\
\hline
USNO B1.0 0707-10151219 & 0.49 & 0.06 \\
\hline
	2a & 0.18 & 0.04  \\
\hline
2MASX J07473002-1917503 & 0.41 & 0.03 \\
\hline
& &  \\
 \multicolumn{3}{|c|}{\textbf{Map M2/Band 7}} \\
 \hline
& &  \\
Source ID & Integrated flux (mJy) & RMS (mJy)\\
\hline
BCG PKS0745-191 & 4.24 & 0.02  \\
\hline
1b & 0.78 & 0.06 \\
\hline
2b & 0.30 & 0.07 \\
\hline
3b & 0.37 & 0.06 \\
\hline
 4b & 0.48 & 0.06 \\
\hline
\hline 
\end{tabular}
\end{center}
\end{table*}
It is impossible to compute an integrated flux over sources 1b+3b as a single extended source because the shape of the resulting source is neither circular nor elliptic. The fluxes measurement are therefore subject to a large uncertainty that depends on the chosen source shape.

The correlation coefficient between maps M1 and M2 is only $2.1 \times 10^{-2}$, hence these maps can be considered statistically independent.  Therefore, they can be combined  
in a map (hereafter M3) where the noise level is reduced, although this does not mean
the improvement of the S/N for each point source. Such operation was carried out by the algorithm Feather implemented in CASA  \citep{mcm07}. This algorithm  performs a weighted addition in the Fourier domain of  M1 and M2
in such a way as to obtain a map with the highest resolution among the two.

The zero-mean unit-variance version of M3 is shown in Fig.~\ref{fig10}(f). Given the different areas covered by M1 and M2, M3  shows an area of size that is similar but
slightly smaller than that of M2. The corresponding histogram $H(x)$ of the pixel values is shown in 
Fig.~\ref{fig11}(c), whereas the two-point correlation function is given
in Fig.~\ref{fig12}(c). In this map $948$ peaks are identified which, as shown by $\rho_p(r)$ in Fig.~\ref{fig13}(c),  can be assumed to be independent and identically distributed with good accuracy. These peaks
provide a maximum likelihood estimate $\hat{\kappa}=0.96$. However, only three of these peaks have a PFA smaller than $10^{-3}$. 
They are highlighted in Fig.~\ref{fig15}(f). 
The SPFA is equal to $3.2 \times 10^{-1}$,  $2.7 \times 10^{-1}$ and $4.0 \times 10^{-3}$ for the detections 1c-3c, respectively. The corresponding probability to have $N_{\rm FD} \leq 1$ is about $0.90$.
As before, if the standard PFA~\eqref{eq:fd1} had been adopted,  the number of detections should have been $326$.  
Two additional sources are detected in M3, while the source 3c is the same source as 3b (or 1b+3b) detected in M2.
The positions and size of the detected source in M3 are reported in the bottom rows of Tab.~\ref{list}. 

The outcome of our investigation implies the following: the sources detected in M2 are not visible in M1 because of the lower spatial resolution in band 3. Source 3c in M3 is the same as source 3b (1b+3b) in M2, while the sources 1c and 2c in M3 are undetected both in M1 and  M2, even if they are visible in M2 at the same spatial position. 
Moreover, contrary to the PFA,  the SPFA permits us to quantify the real risk of a detection claim. Although the PFA of the sources 1b-2b in M2 is $\approx 8.9 \times 10^{-4} $ and $\approx 7.4 \times 10^{-4}$, respectively,
the SPFA indicates that these detections have a confidence level of $43\%$  and  $50\%$ (see also Fig.~\ref{fig16}).

\section{Application to a real ATCA map} \label{sec:atca}

\begin{landscape}
   \begin{figure}
        \resizebox{\hsize}{!}{\includegraphics{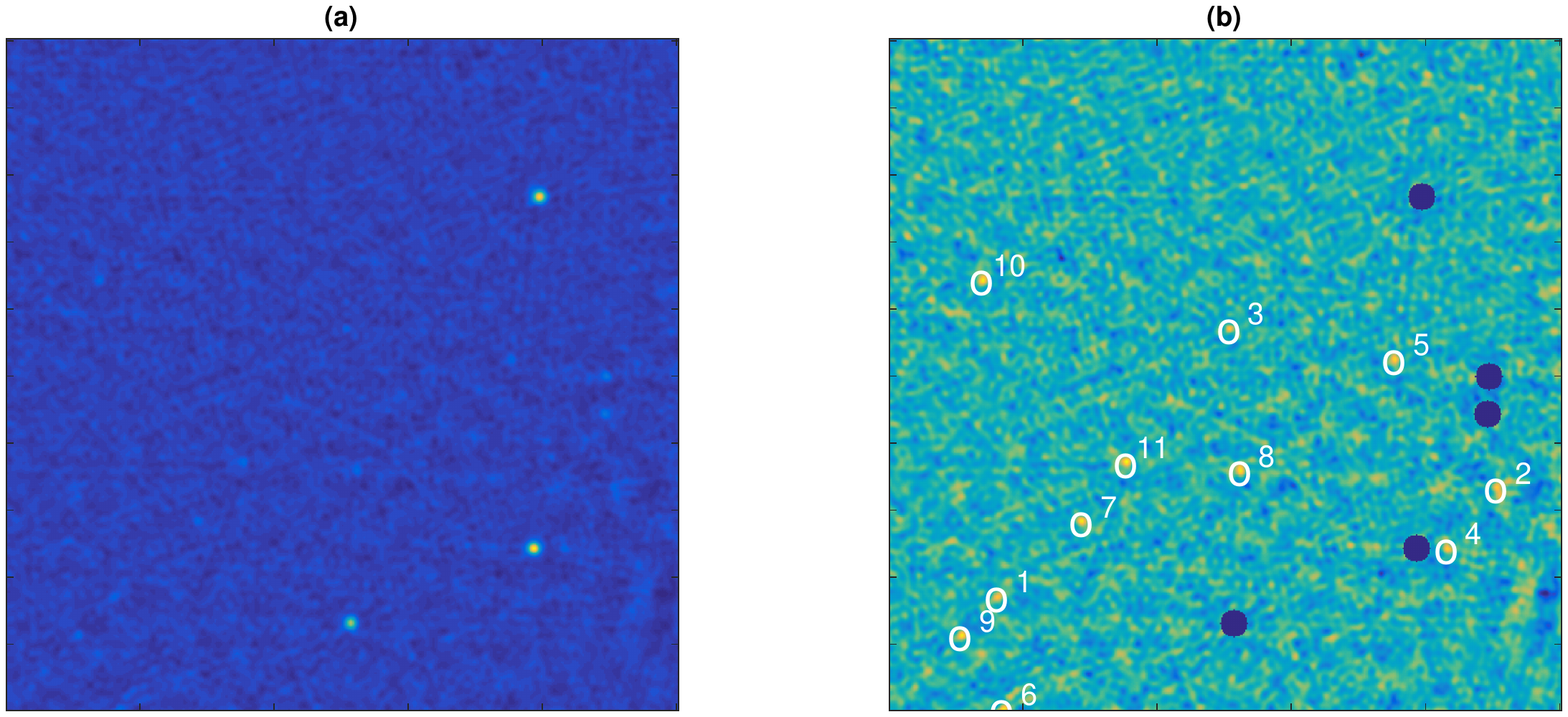}}
        \caption{(a) Original $500 \times 500$ pixels sub-image, standardized to zero mean and unit variance, used for the detection of the point sources; (b) same image with the brightest sources masked (blue circles) and the detected point sources are indicated with white circles and indexed with an increasing number according to the source amplitude.}
        \label{fig17}
    \end{figure}
\end{landscape}

We apply the above method to quantify the detection reliability of point sources extracted from a 500x500 pixels map cropped from a radio image taken with the ATCA array toward the Large Magellanic Cloud at a frequency of $4.8~{\rm GHz}$ by
 \citet{dic05}. This map is shown in Fig.~\ref{fig17}(a). The same map,  standardized to zero mean and unit variance and with some bright sources masked, is shown in Fig.~\ref{fig17}(b).

\begin{landscape}
   \begin{figure}
        \resizebox{\hsize}{!}{\includegraphics{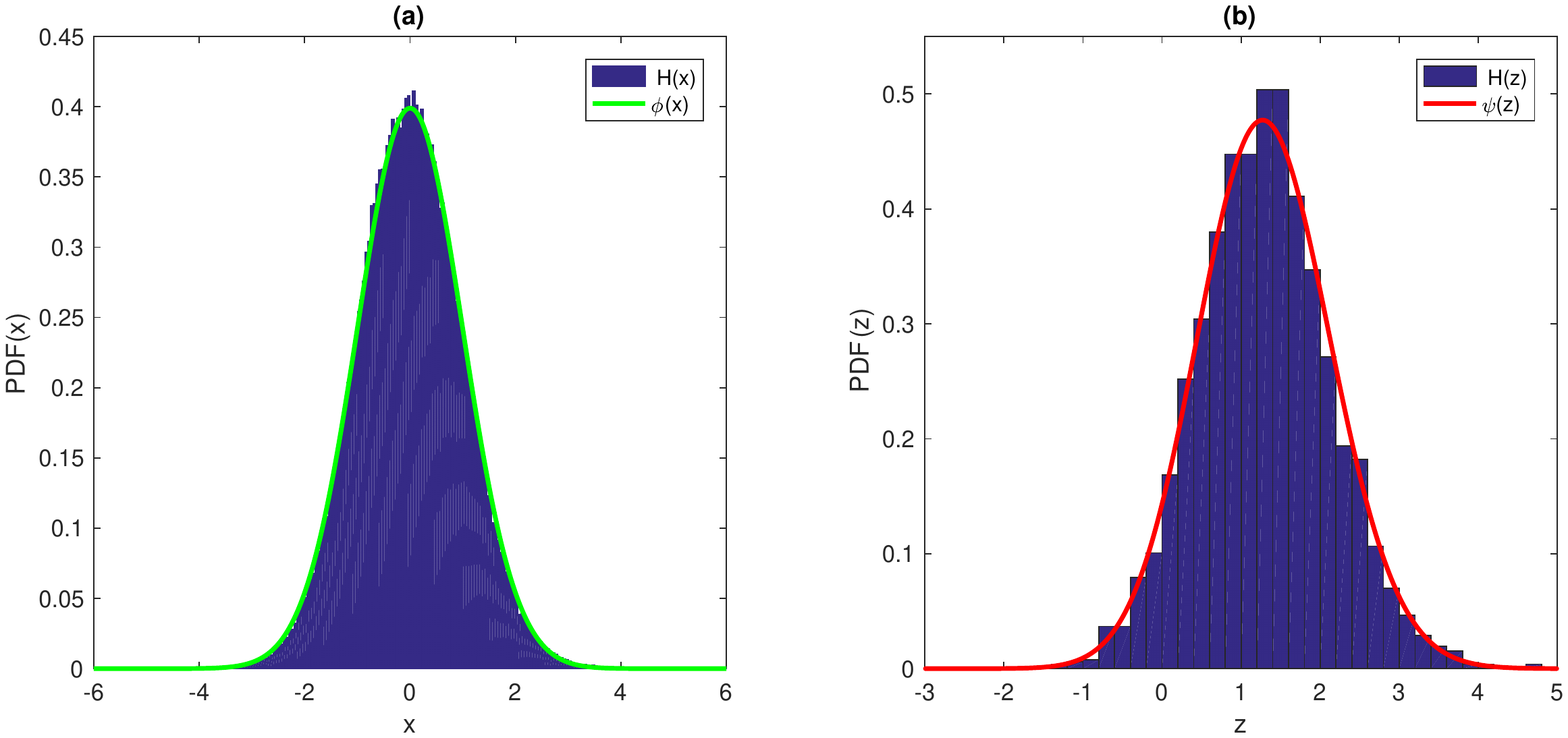}}
        \caption{(a) Histogram $H(x)$ of the pixel values of the map in Fig.~\ref{fig17}(b). For reference, the standard Gaussian PDF $\phi(x)$ is also plotted; (b) histogram $H(z)$ is shown of the values of the peaks in the map
in Fig.~\ref{fig17}(b) vs. the corresponding PDF $\psi(z)$.}
        \label{fig18}
    \end{figure}
\end{landscape}

\begin{landscape}
   \begin{figure}
        \resizebox{\hsize}{!}{\includegraphics{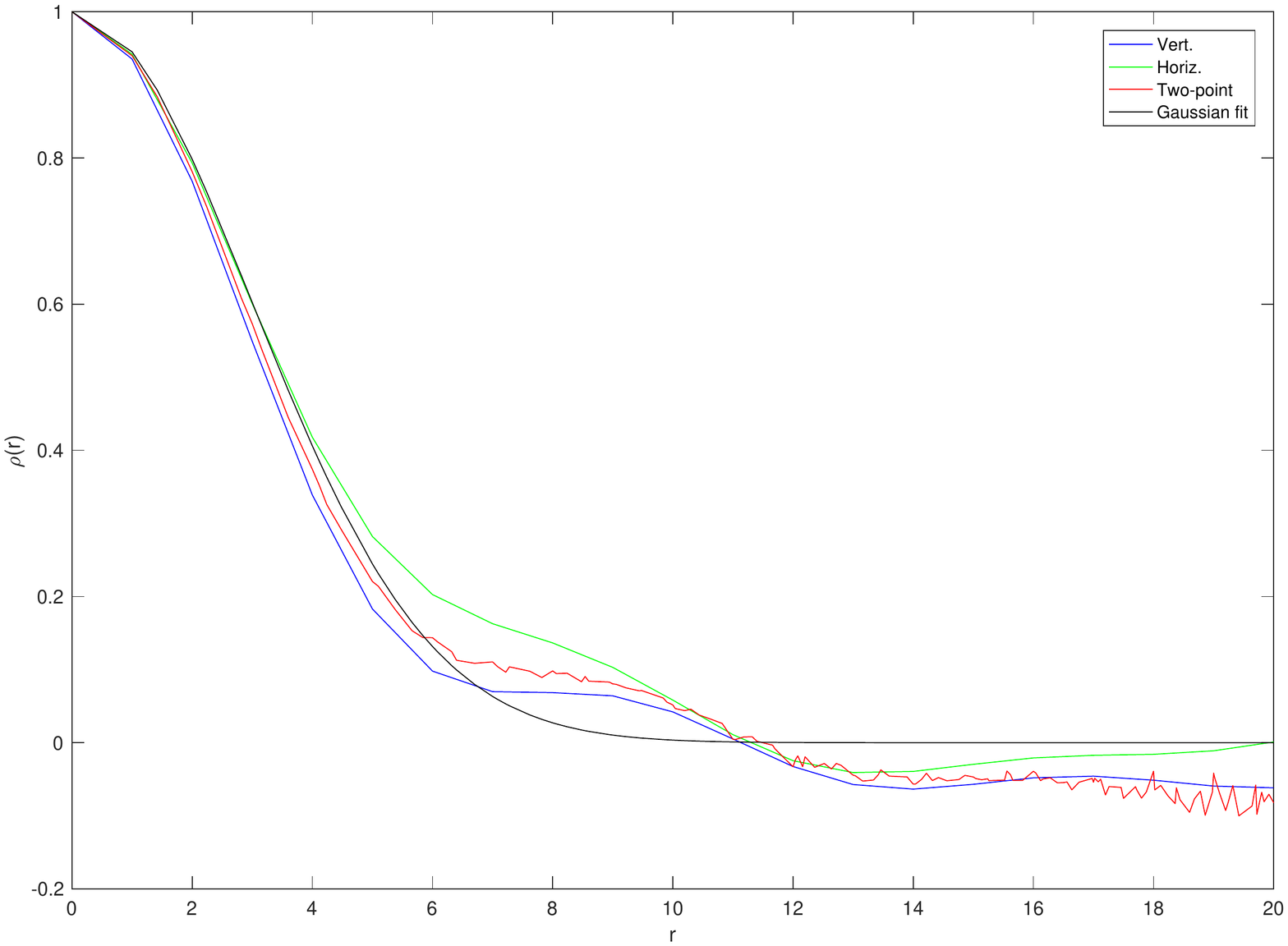}}
        \caption{Autocorrelation function along the vertical and horizontal directions, two-point correlation function, and the corresponding least-squares fit with a Gaussian function of the ATCA map. The inter-pixel distance $r$ is in pixels units.}
        \label{fig19}
    \end{figure}
\end{landscape}

\begin{landscape}
   \begin{figure}
        \resizebox{\hsize}{!}{\includegraphics{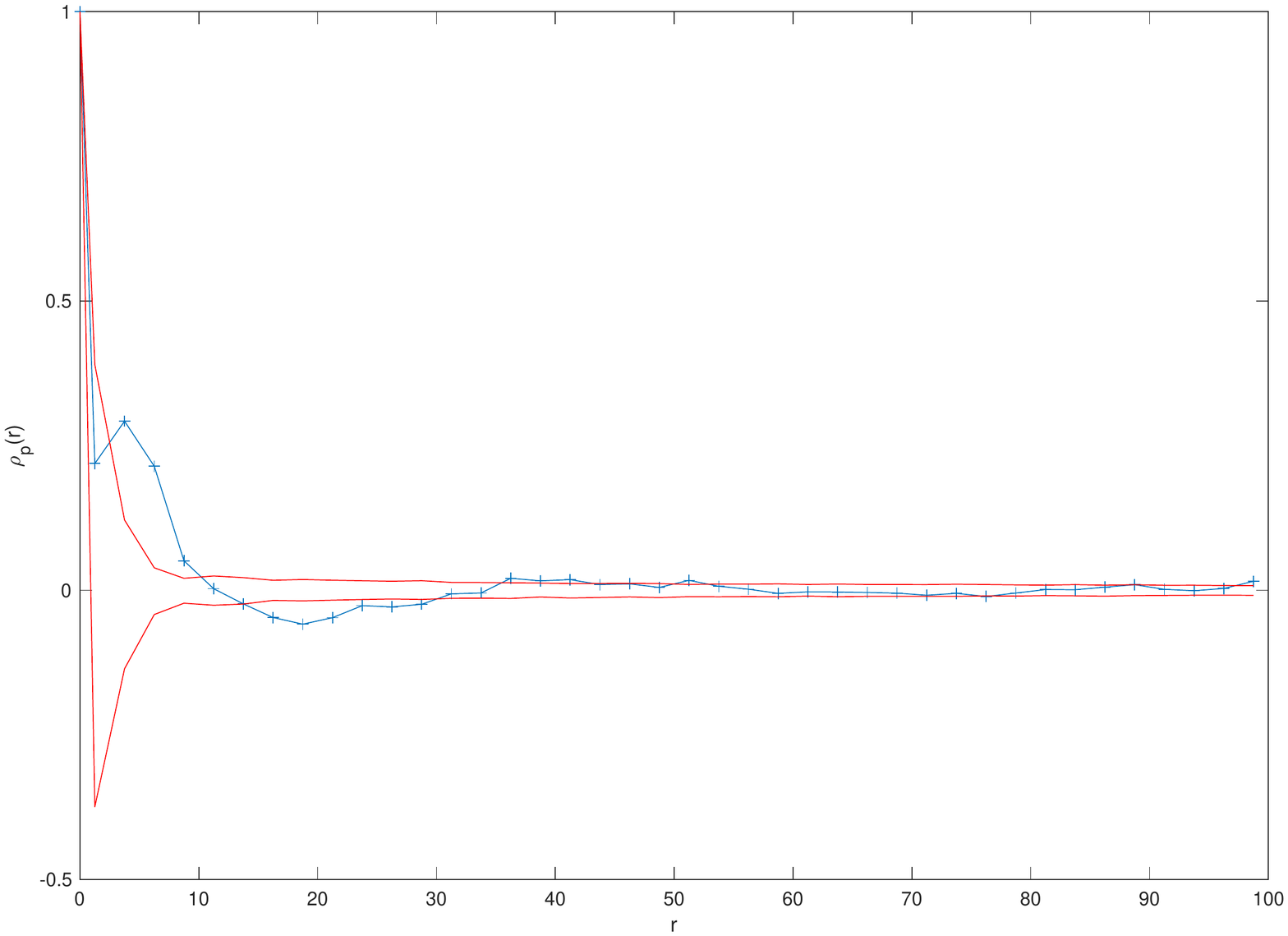}}
        \caption{Two-point correlation function $\rho_p(r)$ for the peaks in the map in fig.~\ref{fig17}(b). The inter-point distance $r$ is in pixels units. The two red lines define the $95\%$ confidence band. These were obtained by means of a bootstrap method based on the $95\%$ percentile envelopes of the two-point correlation functions obtained from $1000$ resampled sets of peaks with the same spatial coordinates as in the original signal, but whose values were randomly permuted.}
        \label{fig20}
    \end{figure}
\end{landscape}

The {\it uv}-plane coverage of this instrument is less homogeneous than that of ALMA (see Fig.~2 in \citet{dic05} for comparison with Fig.~\ref{fig09}), hence the noise background is expected to be less uniform. This is also visible in the map itself, which shows artificial structures introduced by the gridding algorithm.
In spite of this, as Fig.~\ref{fig18}(a) shows, the histogram $H(x)$ of the value of the pixels indicates that the noise is Gaussian. In the map there are $2580$ peaks whose histogram $H(z)$  is shown in Fig.~\ref{fig18}(b). 
The isotropy of the noise background is supported by Fig.~\ref{fig19}, which compares the autocorrelation functions along the vertical and horizontal directions
with the two-point correlations function $\rho(r)$. The two-point correlation function $\rho_p(r)$ of the peaks is shown in Fig.~\ref{fig20}. A weak correlation, probably due to the weighting method applied to the map to fill the gaps in the {\it uv}-plane coverage, is present only for small inter-point distances. 
Hence, also the iid condition for the peaks is approximately satisfied. The estimated  $\hat{\kappa}$ is  $\approx 1$ with the corresponding PDFs $\psi(z)$ shown in Fig.~\ref{fig18}(b).
All these results, together with the good agreement of $H(z)$ with $\psi(z)$, indicate that in spite of the poorer uv-plane coverage of the ATCA map the method can be applied to this map too.

With a threshold $u$  set to $\approx 4.5$, corresponding to a PFA
equal to $1.25 \times 10^{-4}$, $11$ candidate point sources are detected with a SPFA of $0.20$, $0.13$, $0.10$, $0.02$, and $0.01$, respectively, and the remaining with values well below $10^{-3}$. These value correspond
to a small risk of false detection given that that probability of $N_{\rm FD} = 0$ is about $0.6$, which becomes $0.93$ for $N_{\rm FD} \leq 1$.  As a consequence, these peaks are good candidate for a follow-up observations. The sources \#1, \#3, \#6, and \#7 can be identified with the SSTSL2 sources reported by \citet{mau03}. The corresponding identifiers are \object{SUMSS J051432-685446}, \object{SSTSL2 J051230.46-682802.0}, \object{SSTSL2 J051555.19-690746.2}, and 
\object{SSTSL2 J051230.46-682802.0}.

\section{Final remarks} \label{sec:conclusions} 

In this paper we have reconsidered the procedure suggested by \citet{vio16} for the correct computation of the probability of false detection (PFA) of the matched filter (MF) to the case of weak signals with unknown position. In particular, 
we showed that although the PFA is useful for a preliminary selection of the candidate detections, it is not able to quantify the real risk in claiming a specific detection. For this reason we introduced a new quantity called  the 
specific probability of false alarm,  which can provide this kind of information.

We applied this procedure to two ALMA maps at two different frequencies and we highlighted the presence of $7$ potential new point sources (2 in M1, 3 in M2, and 2 in M3). The same procedure applied to an ATCA map provided
$11$ potential point sources.

\begin{acknowledgements}
This research has been supported by a ESO DGDF Grant 2014 and R.V. thanks ESO for hospitality. \\

This paper makes use of the following ALMA data: ADS/JAO.ALMA\#2012.1.00837.S. ALMA is a partnership of ESO (representing its member states), NSF (USA) and NINS (Japan), together with NRC (Canada) and NSC and ASIAA (Taiwan) and KASI (Republic of Korea), in cooperation with the Republic of Chile. The Joint ALMA Observatory is operated by ESO, AUI/NRAO and NAOJ.

\end{acknowledgements}

\listofobjects

\clearpage
\clearpage
\clearpage
\clearpage
\clearpage
\clearpage
\clearpage
\clearpage
\clearpage
\clearpage
\clearpage
\clearpage
\clearpage
\clearpage
\clearpage
\clearpage
\clearpage
\clearpage
\clearpage
\clearpage


\begin{thebibliography}{}
\bibitem[Cheng \& Schwartzman (2015a)]{che15a} Cheng, D., \& Schwartzman, A. 2015a, Extremes, 18, 213
\bibitem[Cheng \& Schwartzman (2015b)]{che15b} Cheng, D., \& Schwartzman, A. 2015b, arXiv:1503.01328 [math.PR]
\bibitem[Cheng \& Schwartzman (2016)]{che16} Cheng, D., \& Schwartzman, A. 2016, arXiv:1511.06835 [math.PR] 
\bibitem[Dickel et al. (2005)]{dic05} Dickel, J., Gruendl, R., McIntyre, V., \& Milne, D.K. 2005, AJ, 129, 790
\bibitem[El-Samie et al. (2013)]{els13} El-Samie, F.E.A., Hadhoud, M.M., \& El-Khamy, S.E. 2013, Image Super-Resolution and Applications (London: CRC Press)
\bibitem[Galatsanos et al. (2005)]{gal05} Galatsanos, N.P., Wernick, M.N., Katsaggelos, A.K., \& Molina, R. 2005, in Handbook of Image \& Video processing, Editor Al Bovik, 203 (New York: Academic Press)
\bibitem[Hogg et al. (2013)]{hog13} Hogg, R.V., McKean, J.W., \& Craig, A.T. 2013, Introduction to Mathematical Statistics (New York: Pearson) 
\bibitem[Hopkins et al. (2002)]{hop02} Hopkins, A.M., Miller, C.J., \& Connolly, A.J. 2002, AJ, 123, 1086 
\bibitem[Katsaggelos et al. (1993)]{kat93} Katsaggelos, A.K., Lay, K.T., \& Galatsanos, N.P. 1993, IEEE Transaction On Image Processing, 2, 417 
\bibitem[Kay (1998)]{kay98} Kay, S.M. 1998, Fundamentals of Statistical Signal Processing: Detection Theory (London: Prentice Hall)
\bibitem[Jain (1989)]{jai89} Jain, A.K. 1989, Fundamentals of Digital Image Processing (London: Prentice Hall)
\bibitem[Lagendijk \& Biemond (1991)]{lag91} Lagendijk, R.L, \& Biemond, J. 1991, Iterative Identification and Restoration of Images (New York: Springer Science+ Business Media) 
\bibitem[Levy (2008)]{lev08} Levy, B.C. 2008, Principles of Signal Detection and Parameter Estimation  (New York: Springer Science + Business Media)
\bibitem[L\'opez-Caniego et al. (2005a)]{lop05a} L\'opez-Caniego, M., Herranz, D., Barreiro, R.B., \& Sanz, J.L. 2005, MNRAS, 2005, 359, 993
\bibitem[L\'opez-Caniego et al. (2005b)]{lop05b} L\'opez-Caniego, M., Herranz, D., Barreiro, R.B., \& Sanz, J.L. 2005, Journal on Applied Signal Processing 2005,15, 2426
\bibitem[Macmillan, \& Creelman (2005)]{mac05}  Macmillan, N.A., \& Creelman, C.D. 2005, Detection Theory: a User's Guide (Mahwah: Lawrence Erlbaum Associates)
\bibitem[Majumdar \& Comtet (2005)]{may05} Majumdar, S.N., \& Comtet, A. 2005, Journal of Statistical Physics, 119, 777
\bibitem[Mauch et al. (2003)]{mau03} Mauch, T., Murphy, T., Buttery, H.J. et al. 2003, MNRAS, 342, 1117
\bibitem[McMullin et al. (2007)]{mcm07} McMullin, J. P., Waters, B., Schiebel, D., Young, W., \& Golap, K. 2007, Astronomical Data Analysis Software and Systems XVI (ASP Conf. Ser. 376), ed. R. A. Shaw, F. Hill, \& D. J. Bell (San Francisco, CA: ASP), 127 
\bibitem[Miller et al. (2001)]{mil01} Miller, J.C., Genovese, C., Nichol, R.C. et al. 2001, AJ, 122, 3492
\bibitem[Tuzlukov (2001)]{tuz01} Tuzlukov, V.P. 2001, Signal Detection Theory (New York: Springer Science + Business Media)
\bibitem[Vio et al. (2002)]{vio02} Vio, R., Tenorio, L., \& Wamsteker, W. 2002, A\&A, 391, 789
\bibitem[Vio et al. (2003)]{vio03} Vio, R. et al. 2003, A\&A, 401, 389
\bibitem[Vio et al. (2004)]{vio04} Vio, R., Andreani, P., \& Wamsteker, W. 2004, A\&A, 414, 17
\bibitem[Vio \& Andreani (2016)]{vio16} Vio, R., \&  Andreani, P. 2016, A\&A, 589, A20
\bibitem[Vogel (2002)]{vog02} Vogel, C.R. 2002, Computational Methods for Inverse Problems (Philadelphia: SIAM)
\end{thebibliography}
\end{document}